\documentclass[noshowpacs,twocolumn,aps,superscriptaddress]{revtex4}

\usepackage{graphicx} % Include figure files
\usepackage{dcolumn}  % Align table columns on decimal point
\usepackage{bm}       % bold math
\usepackage{amssymb}  % Some Maths symbols
\usepackage{amsmath}  % More Maths stuff
\usepackage{setspace}
\usepackage{latexsym}
\usepackage{calc}
\usepackage{color}
\usepackage{shadow}
\usepackage{epsfig}
\usepackage{palatino}

\input epsf
\input rotate

\newcommand{\KS}{{\rm KS}}

\def\br{{\bf r}}
\def\bp{{\bf p}}

\def\bq{{\bf q}}

\def\bea{\begin{eqnarray}}
\def\eea{\end{eqnarray}}
\def\ben{\begin{equation}}
\def\een{\end{equation}}

\def\sss{\scriptscriptstyle\rm}

\def\c{_{\sss C}}

\def\xc{_{\sss XC}}

\def\ext{_{\rm ext}}
\def\ee{_{\sss ee}}

\def\2xc{_{\sss 2XC}}

\def\half{\frac{1}{2}}

\def\bp{{\bf p}}
\def\br{{\bf r}}

\def\bq{{\bf q}}

\def\bei{\begin{itemize}}
\def\eei{\end{itemize}}

\begin{document}

\title{Electron Correlation via Frozen Gaussian Dynamics}

\author{Peter Elliott}
\affiliation{Department of Physics and Astronomy, Hunter College and the City University of New York, 695 Park Avenue, New York, New York 10065, USA}
\author{Neepa T. Maitra}
\affiliation{Department of Physics and Astronomy, Hunter College and the City University of New York, 695 Park Avenue, New York, New York 10065, USA}

%%%%%%%%%%%%%%%%%%%%%%%%%%%%%%%%%%%%%%%%%%%%%%%%%%%%

\begin{abstract}
We investigate the accuracy and efficiency of the semiclassical Frozen
Gaussian method in describing electron dynamics in real time. Model
systems of two soft-Coulomb-interacting electrons are used to study
correlated dynamics under non-perturbative electric fields, as well as
the excitation spectrum. The results show that a recently proposed
method that combines exact-exchange with semiclassical correlation to
propagate the one-body density-matrix holds promise for electron
dynamics in many situations that either wavefunction or
density-functional methods have difficulty describing. The results
also however point out challenges in such a method that need to be
addressed before it can become widely applicable.
\end{abstract}

\maketitle

\section{Introduction}
Accurately capturing the correlated motion of electrons in atoms,
molecules, and solids remains an active research area today.  In
ground-state electronic structure problems, the difficulties of
solving many-electron systems are rolled up into the correlation
energy, a.k.a. {\it the stupidity energy}~\cite{F72}. It has been
examined in many different scenarios and various limits, but its
complexity still haunts us today.
%Perhaps expressing this frustration, Feynmann refers to the
%correlation energy as {\it the stupidity energy}, in his lectures on
%statistical mechanics\cite{F89}.
Essentially representing the deviation of the many-electron wavefunction  from an antisymmetrized product of single-particle orbitals, correlation plays a significant role in time-dependent problems as well.
When atoms and molecules are exposed to either perturbative or
non-perturbative external fields, fascinating and subtle electron
interaction effects mean one must go beyond the ``single active electron'' picture, yet to solve the
time-dependent Schr\"odinger equation  (TDSE) for more than two electrons in
strong fields pushes today's computational limits~\cite{PMDT00,DBK10}.
 
The advent of time-dependent density functional theory (TDDFT) in
1984~\cite{RG84} opened up the possibility of using 
single-particle orbitals to describe the dynamics of interacting
electrons exactly.  In TDDFT a single slater determinant (SSD) in
 a non-interacting system (the Kohn-Sham (KS) system) is propagated in
 time, such that it reproduces the exact time-dependent density of the
 true interacting system. Clearly the KS determinant is not the true interacting wavefunction, nor is it supposed to be an approximation to it; nevertheless by the rigorous theorems of TDDFT all observables of the true, correlated, electronic system can be in principle extracted from it exactly. In practise, approximations are needed for the exchange-correlation functionals,  limiting the accuracy of the method.
TDDFT is a method of choice in the
 linear-response regime where the photo-spectra of atoms, molecules
 and clusters can be found accurately. The
 computational effort involved scales relatively well, so the spectrum of large systems
 such as the green fluorescent protein\cite{MLVC03}, or even candidates for solar cells~\cite{SRVB09}, have been studied.

Although it has proved successful for a wide variety of time-dependent
problems~\cite{EFB07,MUNRB06}, its progress for real-time dynamics in
non-perturbative fields has been slower. Three major obstacles involve
the lack of memory in the commonly-used functional approximations,
lack of good functional approximations to extract observables not
directly related to the density, and the inaccuracy of the usual
functional approximations when the true wavefunction evolves far from
a SSD~\cite{RHGM09,RRM10}.

Recently a method that accounts for electron correlation
semiclassically has been proposed~\cite{RRM10} that will, in principle, remedy the aforementioned problems of
TDDFT. The method operates within the context of time-dependent
density-matrix functional theory (TDDMFT), where the one-body density
matrix is propagated in time.  The second-order density-matrix enters
into the equation for the one-body density-matrix, so is approximated
as a functional of the one-body density-matrix in TDDMFT. However it
is still difficult to find approximations in TDDFMT that work well
when the true wavefunction evolves from being close to a SSD to being
far from one~\cite{GGB10,AG10,RP11}, i.e. those developed so far cannot dynamically change
``occupation numbers''.  In the proposed method of Ref.~\cite{RRM10},
a separate semiclassical propagation is done from which the
correlation potential is extracted, and used to drive the actual
propagation. The approach naturally contains memory effects carried
along by classical trajectories, all one-body observables may be
immediately extracted, and occupation numbers do dynamically evolve.

Whether the proposed method is accurate for a given application
however depends on how close the semiclassical dynamics is to the true
dynamics. In the method, only the correlation component of the
dynamics is treated semiclassically, while all other terms (including
Hartree and exchange), are treated exactly. However, if the
semiclassical dynamics deviates far from the true dynamics, the
proposed method is unlikely to be accurate. Therefore it is worth
investigating how well a semiclassical description of the {\it entire}
dynamics is for the problems of interest.  This is the goal of the
present paper. We compute the semiclassical dynamics (within the Frozen Gaussian approximation) of 
model two-electron systems where the exact dynamics
can be numerically exactly computed, so that we can test the accuracy of the
semiclassical dynamics against this. We can also compare with the
exact KS system: that is, using the exact exchange-correlation potential
(that yields the exact density), but comparing observables such as momentum-densities,
computed directly from the KS SSD (instead of using the unknown
appropriate observable-functionals as in exact TDDFT). Our examples
model the following phenomena: states of double-excitation in spectra,
dynamics in a strong oscillating field, and population transfer to
excited states via resonant driving, or via an optimal field. The
chosen systems are one-dimensional models of a quantum dot and a
helium atom. 

We show both successes and failures of the semiclassical propagation
for these applications.  We expect however that the proposed method of
Ref.~\cite{RRM10} is more accurate than such a semiclassical treatment
of the entire problem since there it is only the correlation component
that is approximated semiclassically. So the results of our studies
are expected to improve once applied in the context of the method of
Ref.~\cite{RRM10}.

Our results also have relevance for semiclassical studies of dynamics
in their own right (i.e. outside of density-functional-type
methods). Although semiclassical treatments of dynamics have seen much
interest for nuclear dynamics (eg. Refs.~\cite{M98,TW04,K05}), there have
only been a few applications to electron
dynamics~\cite{SR99,SR99b,K02,VR04,YGPB04,TT06,HK07} and almost all involved a single electron. 
Part of the reason is that in atoms
there is a significant probability that classical trajectories will
autoionize after a few cycles in the atomic well:
through the electron interaction, one electron gains energy from the
other and ionizes while the other drops below its zero-point
energy; a process that is classically allowed but quantum-mechanically forbidden. This is a manifestation of what has been called the  ``zero-point-energy'' problem~\cite{HK07,MHD89,BGS89}, and happens also when there are coupled vibrational modes, for example.
These trajectories create a lot of noise in the semiclassical
sum, making the convergence very difficult, and so are typically
terminated. For example, in Ref.~\cite{HK07}, accurate spectra of the He atom in both its collinear $Zee$ and $eZe$ configurations were obtained, by discarding trajectories where one electron reaches a certain threshold distance. Theoretically, a
question is whether fundamentally these trajectories should be
excluded from the semiclassical sum. That is, if the semiclassics
could be done with an infinite number of trajectories, whether these
trajectories incorrectly contribute to the semiclassical sum. We find
that indeed their contribution to the semiclassical
sum decreases as more trajectories are added:
in this sense, they are valid classical trajectories, and via
phase-interference, the semiclassical dynamics restores the
zero-point-energy, eliminating their contribution from the
semiclassical sum. In practise, nevertheless, we must deal with a
finite number of trajectories, so sensible ways to discard these trajectories
must be devised. 
%We come back to this in Section~\ref{sec:sftcAI}. 
Our results on the model atoms, as well as those on a model quantum dot, lend support to the conclusions of Ref.~\cite{HK07} that semiclassical dynamics is useful for electrons, for both spectra and dynamics in non-perturbative fields (in some cases). A draw of semiclassics is the interpretive power it carries and we hope to use it also to 
to interpret mechanisms for processes involving interacting electrons, e.g. the role of correlation in multiphoton ionization and multielectron ionization. 

The  ``flavor'' of semiclassics explored in the
current work is the Frozen Gaussian (FG) dynamics, originally proposed
by Heller~\cite{H81}. This is not one of the rigorous semiclassical
propagators that satisfy the TDSE up to order $\hbar$; it is however
closely related to the rigorous Heller-Herman-Kluk-Kay (HHKK)
propagator~\cite{HK84,KHD86,TW04,K05,M98,GX98}, which is based on a coherent-state
representation of the semiclassical
propagator. In Section~\ref{sec:Background}, 
after a brief review of TDDFT and TDDMFT that explains the motivation behind Ref.~\cite{RRM10},
we review some background regarding FG dynamics and how it is
proposed to be used in the method of Ref.~\cite{RRM10} in
density-matrix propagation. Section~\ref{sec:results} presents results on one-dimensional model systems of two electrons, using FG for dynamics and spectra. Finally, we make some conclusions in Sec.~\ref{sec:Conclusions}.

\section{Background}
\label{sec:Background}
In this section, we briefly review TDDFT,  the motivation behind going from TDDFT to a
density-matrix approach, why we consider a
semiclassical approach to correlation, and semiclassical dynamics. 

\subsection{TDDFT and its challenges}
\label{sec:BackgroundTDDFT}
TDDFT is an exact reformulation of the non-relativistic time-dependent
quantum mechanics of many-body systems~\cite{RG84,MUNRB06}.  It
operates by mapping the true problem of interacting electrons into a
ficitious, non-interacting system of fermions, called the Kohn-Sham
(KS) system, whose one-body density is, in principle, exactly that of
the true system, and from which, in principle, all properties of the
true system can be obtained. The many-body effects of the interacting
system are modelled by a one-body potential, called the
exchange-correlation potential $v\xc$.
%one-body potential in which the non-interacting fermions live, called the KS potential, $v\s$, is usually decomposed into three parts: the external potential $v\ext$ which is the potential applied to the true interacting system of electrons (eg from nuclear attraction and laser fields), the  classical Hartree potential, $v\H$, and the exchange-correlation potential $v\xc$ which is a one-body potential accounting for many-body interaction effects.
Similar thus in flavor to ground-state density functional theory, its
functionals are nevertheless quite different. For example, $v\xc$ functionally depends on the entire history of the
density, as well as the initial-state of the true system and the
initial state of the KS system,
$v\xc[n,\Psi_0,\Phi_0](\br,t)$\cite{M06}. This {\it memory-dependence} is
however lacking in most of the applications today: they use an
``adiabatic'' approximation, which simply feeds the time-evolving
density into a ground-state approximation:
\ben
v\xc^{\rm adia} [n,\Psi_0,\Phi_0](\br t) = v\xc^{\rm gs} [n] (\br)|_{n(\br)=n(\br t)}\;.
\label{adia}
\een
With this simple approximation, TDDFT has had great success in
calculating spectra and response: computationally it scales similar to
methods such as time-dependent Hartree-Fock (TDHF) or
configuration-interaction singles, but with accuracy that is usually
far greater.  Yet, this simple approximation is also behind why  in some  cases approximate TDDFT fails: e.g. to capture states of 
double-excitation character a frequency-dependent
kernel is required~\cite{TH00,MZCB04,EGCM11}, but the adiabatic approximation
 yields a frequency-independent kernel.

The lack of memory in the usual functional approximations is also one
of the reasons why the application of TDDFT to real-time dynamics
applications has not progressed so fast. Model systems for which exact
results are available indicate that memory-dependence can sometimes be
a crucial factor in the exchange-correlation
potential~\cite{HMB02,MB01,MBW02,WU08,U06}.  Another difficulty in
real-time dynamics, is the problem of ``observable-functionals'':
although in principle all properties of interest can be extracted from
the KS system, it is not known {\it how} to extract those that are not
directly related to the density.  One may write these observables as
operators on the exact wavefunction, however substituting the KS
wavefunction is usually a poor approximation. For example, for
momentum distributions, it was shown\cite{RHGM09} for a model system
with one electron ionizing that the KS momentum distribution incorrectly develops
strong oscillations as the electron moves away. In the case of
ion-momentum-recoil upon ionization of a model He atom, the KS
momentum-distributions were found to be drastically wrong, displaying
a single maximum instead of the characteristic two hump structure, and
with a significantly overestimated
magnitude\cite{WB07}. Double-ionization probabilities are another
example where a more sophisticated observable functional is
needed~\cite{WB06}.

A third problem arises when the true wavefunction evolves far from an
SSD~\cite{MBW02,RRM10}.  An illustration of this is in certain quantum control problems,
where a laser pulse is found to take the system to a specified target
e.g. populating an excited state. For example, a pulse can be found to
move the $1s^2$ ground state of Helium to the $1s2p$ excited
state. As TDDFT stays in a SSD, with a single
doubly-occupied orbital, for all times, it has great difficulty in
describing a state that fundamentally is best described by two
SSDs. It should be emphasized here that in principle, TDDFT can describe
this situation, however the exchange-correlation potential and observable-functionals are
extremely difficult to approximate.

%so one
%has the time-dependent KS equation: \ben
%\label{TDKS}
%i \frac{\partial\phi_{j} (\br t)}{\partial t}  = \left( - \frac{\nabla^{2}}{2} + v\s[n](\br t) \right)\phi_{j}(\br t) 
%\een
%where the time-dependent density of the KS system matches exactly that
%of the interacting system. As opposed to the ground-state case, the KS
%potential is not given by a functional derivative and is a much more
%difficult object to approximate. Moreover, the potential depends on
%the either history of the density at all previous times, as well as
%initial interacting and non-interacting wavefunctions,
%$v\s[n,\Psi_0,\Phi_0](\br,t)$\cite{M06}

\subsection{Time-Dependent Density Matrix Functional Theory (TDDMFT)}
To overcome some of the difficulties of TDDFT, we may attempt to use
the one-body reduced density matrix, defined below, as the fundamental
variable. This has the advantage of containing more information than
the density while retaining similar concepts and system-size scaling as TDDFT.
 An advantage is that
functionals for the momentum distribution and kinetic energy are given exactly in terms
of the density matrix.  
The first-order spin-summed reduced density matrix is
defined as:
\bea
\nonumber
\rho(\br',\br,t)= N\sum_{\sigma_1..\sigma_N}\int d^3r_2..d^3r_N\\
\Psi^*(\br'\sigma_1,\br_2\sigma_2...\br_N\sigma_N,t)\Psi(\br\sigma_1,\br_2\sigma_2...\br_N\sigma_N,t)
%\rho(\br',\br;t) = N\int d\br_2 ...d\br_N \Psi^*(\br',\br_2,...,\br_N;t)\Psi(\br,\br_2,...,\br_N;t)
\eea
and can be diagonalized by the so-called natural orbitals, $\xi_j(\br,t)$:
\ben
\rho(\br',\br;t) = \sum_j \eta_j(t) \xi_j^*(\br',t)\xi_j(\br,t)
\een
here $0 \le \eta_j(t)\le 2$ is the time-dependent occupation number
of the $j$th natural orbital (NO), and $\sum_j \eta(t) = N$. We can
now interpret the quantum control example discussed at the end of Sec.~\ref{sec:BackgroundTDDFT}  as an
issue with the NO numbers. The NOs of the true system will begin with
one NO occupation near $2$ and the rest close to zero, and will end
with two occupation numbers close to $1$. The KS system, beginning in the ground-state SSD, has only
a single natural orbital which is doubly occupied and time-evolution in the one-body KS Hamiltonian means that the NO occupation number always remains at
$2$. By using the density matrix of the true system as basic variable instead, we allow the NO
occupations to change and thus it should be better than KS at capturing
the correct behavior with simpler functional approximations. 

The equation of motion for the density matrix is known as part of the BBKGY hierarchy of equations: 
\bea
i\dot{\rho}(\br',\br;t) &=& \left( -\frac{\nabla^2}{2} + v\ext(\br;t) +\frac{\nabla'^2}{2} - v\ext(\br';t) \right)\rho(\br',\br;t) \nonumber \\
& &+ \int d\br_2 f\ee(\br',\br,\br_2)\rho_2(\br',\br_2,\br,\br_2;t)
\label{eq:rho1dot}
\eea
where $f\ee(\br'\br,\br_2) = 1/|\br-\br_2| - 1/|\br'-\br_2|$ and $\rho_2$ is the second-order spin-summed reduced density matrix, conveniently decomposed as
\bea
\label{rhoc}
\rho_2(\br',\br_2,\br,\br_2;t) &=& n(\br_2)\rho(\br',\br;t) - \rho(\br',\br_2;t)\rho(\br_2,\br;t)/2  \nonumber \\
& & + \rho\c(\br',\br_2,\br,\br_2;t)\;.
\eea
for a closed-shell spin-saturated system. Atomic units ($e^2 = \hbar = m_e = 1$) are used throughout this paper.
Here the terms play the role of Hartree, exchange, and correlation
respectively. Setting $\rho\c=0$ results in the TDHF equations. One could look to
close this expression by expressing $\rho\c$ as a functional of $\rho$
and this is the usual approach in TDDMFT. One naturally seeks an adiabatic approximation in which the time-dependent density-matrix is fed into a 
$\rho_2$-functional bootstrapped from ground-state density-matrix functional theory~\cite{PGB07,PGGB07,GBG08,RP10}  Unfortunately, neither TDHF nor the
adiabatic functionals lead to NO
occupation numbers changing in time~\cite{GGB10,AG10,RP11}. 
Thus we must look further afield to try and fix this.

A possible solution to this was proposed in Ref.~\cite{RRM10} where
$\rho\c$ is given by an auxiliary semiclassical  dynamics
calculation running parallel to the density-matrix evolution. That is,
we time-evolve $\rho$ using Eq.~\ref{eq:rho1dot} with
Eq.~\ref{rhoc}, where all terms are treated exactly
quantum-mechanically {\it except} the last term $\rho\c$ of
Eq.~\ref{rhoc}. This last term is treated as a driving term in this
equation, its value obtained from the parallel semiclassical propagation of the system.
At each time, the second-order density-matrix, the density matrix, and the density are calculated in the semiclassical system, which then uses Eq.~\ref{rhoc} to extract a semiclassical $\rho\c$. This term is then inserted in Eq.~\ref{eq:rho1dot} as a driving
term. In Ref.~\cite{RRM10}, it was argued that such an approach
addresses the main obstacles TDDFT and adiabatic TDDMFT approaches for
real-time dynamics: memory is naturally carried along by the classical
trajectories, initial-state dependence is automatically accounted for,
all one-body observables may be obtained without the need for further
observable-functionals, and NO occupation numbers are time-evolving. We can expect this method to most improve upon the purely semiclassical calculation when in the high-density limit, as the dominating exchange affects will be treated exactly.

Various candidates for the semiclassical method to use in this scheme are possible, with their
numerical efficiency inversely related to their semiclassical rigor
and accuracy~\cite{RRM10}. Most efficient and least accurate is to use
a quasiclassical propagation of the one-body Wigner
function~\cite{H76,BH81}, while least efficient but most rigorous, solving TDSE  exactly to order $\hbar$, is to use the Heller-Hermann-Kluk-Kay (HHKK)
propagator~\cite{HK84,KHD86,TW04,K05,M98,GX98}. 
The
Frozen
Gaussian (FG) method~\cite{H81} may be viewed as an approximation to HHKK, and was argued to be a good candidate for use in a ``forward-backward'' scheme for $\rho_c$~\cite{RRM10}. 
We focus in this paper therefore on FG dynamics, and our present goal is to
investigate how accurate it alone is for electron dynamics. Although
in the scheme of Ref.~\cite{RRM10}, semiclassics is used only for the
correlation component of the dynamics, the scheme is likely to be most
accurate when the semiclassical propagation of the whole system is reasonably
accurate. The purpose of the studies in Section~\ref{sec:results} is therefore 
to study how accurate FG dynamics on model systems is.

\subsection{Frozen Gaussian Dynamics}
\label{sec:BackgroundFG}
Since the earliest days of quantum mechanics semiclassical methods
have been explored, for the purpose of interpreting and understanding
quantum mechanics via the more intuitive classical dynamics, and  also for the purpose of approximation.  These methods construct
an approximate quantum propagator utilizing classical trajectory
information alone. Although there are a variety of forms
(e.g.~\cite{H81,H91,M98,M70,M74,V28,G67,S81,TW04,GX98}), their essential
structure is a sum over classical trajectories:
$
\sum_{\rm cl. traj.} C_i(t)e^{iS_i(t)/\hbar}
$
where $S_i(t)$ is the classical action along the $i$th trajectory, and
the prefactor $C_i(t)$ captures fluctuations around the classical
path. Miller~\cite{M74} showed the
equivalence of different semiclassical representations within
stationary-phase evaluation of the transformations. Semiclassical formulae have been derived both from largely
intuitive arguments~(e.g. Heller's frozen and thawed gaussians~Ref.\cite{H81,H81b}) as well as from rigorous
asymptotic analyses of the quantum propagator~(see
e.g.~Refs.\cite{S81,K05,V28,G67}) that satisfy TDSE to order $\hbar$.  The latter are based on taking
the semiclassical limit of Feynman's path integral for
exact quantum dynamics. This yields a propagator of van Vleck
form~\cite{V28}, that requires solving a boundary value problem to
find the classical paths eg. from $x'$ to $x$ in time $t$;
transforming these to ``initial-value'' representations where instead
a sum over an initial coordinate-momentum phase-space is performed,
makes the numerics significantly more feasible, especially for longer
times and more degrees of freedom.  

Semiclassical dynamics captures quantum effects such as interference,
zero-point energy, tunneling (to some extent), while generally scaling
favorably with the number of degrees of freedom.  As the propagator is
constructed from classical trajectories, intuition about the physical
mechanisms underlying the dynamics can be gained.  Although mostly
used for nuclear dynamics in molecules, and condensed phases, there
have been a handful of applications to
electrons~\cite{SR99,SR99b,K02,VR04,YGPB04,TT06,HK07}.  The reasons for the
reluctance of applying semiclassics to electrons were recently
discussed in Ref.~\cite{HK07}. One is that the classical dynamics of
interacting electrons are typically mixed with regular and chaotic
regions. This makes the semiclassical sum over trajectories
increasingly difficult to converge as time evolves. Another is the
classical autoionization problem discussed in the introduction; we
return to this in Sec.~\ref{sec:sftcAI}. A third problem is that the
classical equations of motion become singular at the nucleus, where
the potential diverges Coulombically, requiring the need for
regularization methods to be applied. As Ref.~\cite{SR99} showed
however, this is apparently only a problem when systems are treated in
reduced dimensionality: in three-dimensions, no special techniques are
required to deal with the Coulomb potential except for a set of
measure zero trajectories which hit the nucleus head-on (and these
trajectories can be safely discarded).  Ref.~\cite{HK07} showed that
all these obstacles can be overcome to make semiclassical calculations
of atoms and molecules quite practical.

The FG propagation we are exploring can be expressed mathematically as a simplified version of the 
HHKK  propagator~\cite{HK84,KHD86,TW04,K05,M98,GX98} where 
the $N$-particle wavefunction at time $t$ as a function of the $3N$ coordinates we denote ${\underline{\underline{\br}}}$, is:
\ben
\label{frozg}
\Psi^{\rm FG}({\underline{\underline{\br}}},t) = \int\frac{d\bq_0d\bp_0}{(2\pi\hbar)^N}\langle{\underline{\underline{\br}}}\vert\bq_t\bp_t\rangle e^{iS_t/\hbar}\langle\bq_0\bp_0 \vert\Psi_i\rangle
\een
where $\{\bq_t,\bp_t\}$ are classical phase-space trajectories in
$6N$-dimensional phase-space, starting from initial points
$\{\bq_0,\bp_0\}$, and $\Psi_i $ is the initial state. In
Eq.~\ref{frozg}, $\langle{\underline{\underline{\br}}}\vert\bq\bp\rangle$ denotes the
coherent state:
\ben
\langle{\underline{\underline{\br}}}\vert\bq\bp\rangle = \prod_{j=1}^{3N}\left(\frac{\gamma_j}{\pi}\right)^{1/4}e^{-\frac{\gamma_j}{2}(r_j-q_j)^2 + ip_j(r_j-q_j)/\hbar}
\een
where $\gamma_j$ is a chosen width parameter and $S_t$ is the
classical action along the trajectory $\{\bq_t,\bp_t\}$. In HHKK, each
trajectory in the integrand is weighted by a complex pre-factor based
on the monodromy (stability) matrix.  The pre-factor is time-consuming
to compute, and scales cubically with the number of degrees of
freedom, but in the FG approximation, it is set to unity. As a
consequence, although HHKK solves TDSE exactly to order $\hbar$, the
FG propagator does not; also, although it can be shown that HHKK
results are independent of the choice of width parameter $\gamma_j$,
FG results are not.  For all our calculations we take $\gamma_j=1$.
Neither the HHKK propagation nor FG are unitary; typically we find the
norm of the FG wavefunction decreases with time, and so we renormalize
at every time-step.  (In fact for the typically chaotic dynamics of
interacting systems, the HHKK prefactor, and norm, grows with time~\cite{HK07}).

The phase-space integral in Eq.~\ref{frozg} is done by Monte Carlo,
with the distribution of initial phase-space points weighted according
to the initial distribution, $\vert \langle\bq_0\bp_0 |\Psi_i\rangle
\vert^2$. Although Monte-Carlo methods scale as $\sqrt{N}$ for
positive integrands, the oscillatory phase can make the FG propagation
very difficult to converge. The scheme of Ref.~\cite{RRM10} takes
advantage of the ``forward-backward'' nature of the propagation of the
second-order density-matrix (i.e in the second-order density matrix,
there is both a $\Psi(t)$ and a $\Psi^*(t)$), which leads to some
cancellation of phase for more than two electrons. We also observe
that the spatial-symmetry of the initial-state is preserved during the
evolution (since the Hamiltonian is for identical particles,
exchanging coordinate-momentum pairs of two electrons does not change
the action). As we study here two-electron singlet states, the
wavefunction is spatially-symmetric under exchange of particles.

\section{Results+discussion}
\label{sec:results}
We study the accuracy of FG propagation for several different systems
involving two soft-Coulomb-interacting electrons in a spin-singlet in
one-dimension.  
Because the TDSE for two interacting electrons in one-dimension can be mapped onto a TDSE for a single electron in two dimensions, the exact problem is easily numerically solvable, and we can compare our FG results to exact ones. The exact ones are obtained either using the approximate enforced time-reversal symmetry (aerts) algorithm coded in the octopus code~\cite{octopus}, or our own exponential mid-point rule code (both use a fourth order Taylor expansion for the exponential). 
We also comment on how these compare with
results using various common approximations in TDDFT for either the
exchange-correlation functional (Sec.~\ref{sec:Hookespectra}), or for
observable-functionals (Secs.~\ref{sec:HookeNOpdist} and
\ref{sec:sftctrap}).

In the latter case, we compute observables from usual operators acting directly on the exact KS wavefunction: this
enables us to isolate errors arising from the observable-functional
approximation alone (i.e. as opposed to errors from the approximation used for the exchange-correlation potential).  For two electrons in a spin-singlet, given the
exact time-dependent density, it is straightforward to obtain the
exact KS wavefunction~\cite{HMB02,RPB06,TGK08,LK05}. We briefly
review how.  In one-dimension, the continuity equation reads
\ben
\frac{\partial n(x,t)}{\partial t} + \frac{\partial j(x,t)}{\partial x} = 0 \;,
\label{eq:ctuity}
\een
and since the exact KS system reproduces the exact density at all times, 
\ben
j(x,t) = j_\KS(x,t)= 2 \ \mathfrak{Re} \ \phi_\KS^*(x,t)\frac{1}{i}\frac{\partial\phi_\KS(x,t)}{\partial x} \;.
\label{eq:j}
\een
The last equality follows from the fact that for two electrons in a spin-singlet, there is just one spatial KS orbital $\phi_\KS$ (doubly-occupied).
%\ben
%j_{KS}(x,t) = \frac{1}{i}  \left( \phi_{KS}^*(x,t)\frac{\partial\phi_{KS}(x,t)}{\partial x} - \phi_{KS}(x,t)\frac{\partial\phi_{KS}^*(x,t)}{\partial x} \right)
%\een
Writing $\phi_\KS$ as an amplitude times a phase, we deduce, 
\ben
\phi_{KS}(x,t) = \sqrt{\frac{n(x,t)}{2}}\exp\left(i \int^x \frac{j(x',t)}{n(x',t)} dx'\right).
\label{eq:phiks}
\een
%where, from Eq.~\ref{eq:j}, we deduce
%\ben
%A(x,t) = \int^x \frac{j(x',t)}{n(x',t)} dx' \;.
%\een
So, the procedure is as follows: we solve the TDSE exactly
numerically, finding the exact two-electron correlated
wavefunction. The density and current are then extracted from
$n(x,t) = 2\vert\phi_{KS}(x,t)\vert^2$ and Eq.~(\ref{eq:ctuity}), and then inserted into
Eqs.~(\ref{eq:phiks}). Usual approximations for observables in TDDFT
evaluate the operator corresponding to the observable directly on the
KS state; although this is exact for observables directly
related to the density, there are errors for other observables,
e.g. the momentum-density, which is what we shall look at in examples~\ref{sec:HookeNOpdist} and~\ref{sec:sftctrap}.

In all cases, we will utilize the exact ground-state wavefunction to construct our initial state for propagation. This is to separate out any errors due to a poor ground-state. For a general case where the initial wavefunction is unknown exactly, an appropriate approximate initial state (e.g. sum of a few KS SSD's, or from a high-level wavefunction calculation) would be used.

\subsection{Hooke Model: Spectra}
\label{sec:Hookespectra}
We begin with the Hooke model for a one-dimensional two-electron quantum dot defined by the Hamiltonian
\ben
\hat{H}_0 = \sum_{i=1}^2\left(-\half\frac{d^2}{dx_i^2}+\half x_i^2\right) + \frac{1}{\sqrt{(x_1-x_2)^2+1}}
\een
The Hamiltonian becomes separable when written in relative $r=x_1-x_2$
and center of mass $R = (x_1+x_2)/2$ coordinates, and we use the
quantum numbers of these systems to label the excitations of the full
system. In Table~\ref{t:hooke}, we give the frequencies of the lowest
$8$ singlet excitations of this system, along with labels in
$R,r$. The $r$ label must be even in order to have the correct spatial
symmetry for a singlet state, while even/odd values for $R$ determine
the parity of the state. We have grouped the levels into multiplets whose members become degenerate if the electron-interaction is turned off. Each of the two states in the second and third multiplets are states of double-excitation character: they are mixtures of a single excitation (where, in a non-interacting reference, only one electron is promoted from its ground-state orbital to an excited one), and a double-excitation, where both the electrons occupy excited orbitals~\cite{MZCB04,EGCM11}. The three states in the fourth multiplet are mixtures of a single-excitation and two double-excitations. 

Such states of double-excitation character are well-recognized to be
poorly described with the adiabatic approximation in TDDFT, which is
used in almost all calculations~\cite{TH00,MZCB04}: instead of
producing a multiplet of states, only one state is obtained in the
adiabatic approximation, and it has only single-excitation character.
In order to find the correct frequencies, TDDFT requires a
frequency-dependent kernel to mix the KS single and
double-excitations, as was derived in Ref.~\cite{MZCB04}. The kernel
is an {\it a posteriori} correction to an adiabatic approximation and
although it has been recently tested on a variety of
systems~\cite{CZMB04,MW09,MMWA11,HIRC11}, it is not widely used; almost
all codes still rely on the adiabatic approximation that fails to
capture these.  This is the motivation behind our first exploration of
the FG dynamics: is semiclassical correlation able to capture states
of mixed single and double-excitation character? The answer, as shall
see shortly, is yes, at least approximately.

\begin{table}
\caption{The singlet excitation frequencies $\omega_n=E_n-E_0$, where the ground-state energy $E_0 = 1.774040$ a.u., for the Hooke model. For each excitation, the energy can be written $E_n = E_R+E_r$, where $R$,$r$ are labels in the center of mass and relative coordinates respectively.}
\label{t:hooke}
\begin{ruledtabular}
\begin{tabular}{r|rr}
Label (R,r)	& $\omega_{n}$ & $\omega^{FG}_{n}$ 	\\
\hline
1,0				& 1.000000 & 1.0	\\
\\
0,2				& 1.734522 & 1.6	\\
2,0				& 2.000000 & 2.0	\\
\\
1,2			 	& 2.734522 & 2.6	\\
3,0				& 3.000000 & 3.0	\\
\\
0,4				& 3.648334 & 3.5	\\
2,2				& 3.734522 & 3.6	\\
4,0				& 4.000000 & 4.0	\\
\end{tabular}
\end{ruledtabular}
\end{table}

We compute the FG spectra by Fourier-transforming a time-propagation. It is common to start in a linear {\it kicked} state defined as 
\ben
\Psi_k(x_1,x_2) = e^{ik(x_1+x_2)}\Psi_0(x_1,x_2)
\een  
where $\Psi_0$ is the exact ground-state wavefunction, in order to populate the excited states and then to Fourier transform the time-dependent dipole moment, $d(t)$, to find the dipole-spectra:
\ben
d(\omega) = \int dt e^{i\omega t}d(t)
\een
The {\it kick} is equivalent to applying an impulse electric field,
$\delta v\ext(x,t) = k\delta(t)x$,  to the system\cite{YB96,YNIB06,octopus,TK09}. For The spectra
found this way consists of peaks at the excitation frequencies, the
widths of which becomes smaller the longer you propagate the system.
For Hooke's model, this method needs a slight adjustment. In linear response, the {\it kick} yields a perturbation that is proportional to $R$,
which only couples  the ground state $(0,0)$ to the $(1,0)$ state, as the Hamiltonian is simply a
harmonic oscillator in this coordinate; the harmonic potential has the special property that a dipole can only connect two states differing by one quantum number. This can also be seen as a
consequence of the Harmonic Potential Theorem\cite{D94}. So in the
exact system, we only see a single peak in the dipole moment when using a linear kick. In order to see higher excitations, we use
quadratic and cubic kicks and plot  the quadrupole and third-order
moments respectively. For example the quadratic kick, $e^{i
  k(x^2+y^2)}$, in the linear-response regime will be proportional to
both $R^2$ and $r^2$, which, using the symmetry and parity discussion
earlier, will couple most strongly to the $(2,0)$ and $(0,2)$ states.

\begin{figure}[tb]
\unitlength1cm
\begin{picture}(12,6.2)
\put(-6.2,-3.6){\makebox(12,6.2){
\includegraphics{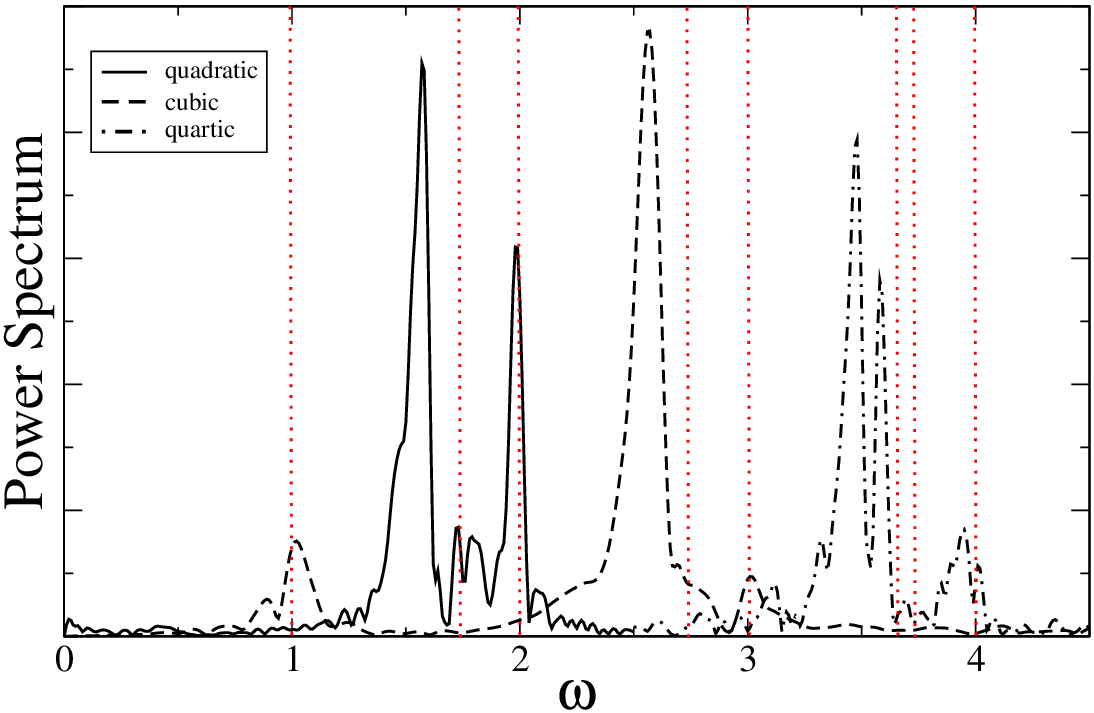}
}}
\end{picture}
\caption{The power spectra of the Frozen Gaussian quadratic, cubic, and quartic moments for the Hooke's model dot. The positions of the exact frequencies are shown as the vertical dashed lines.}
\label{f:hookqspec}
\end{figure}

We begin the FG calculation in the quadratically {\it kicked} state
(using the exact ground state) with $k=0.01$ and use $120000$
classical trajectories. These trajectories are propagated using the
standard leapfrog verlet algorithm for a total propagation time of
$T=200$ a.u with the dipole moment calculated every $0.1$ a.u. The
power spectrum (absolute value squared) of the quadrupole moment is
shown as the solid line in Fig. \ref{f:hookqspec}; we use the power spectrum as it is
usually more stable than either the purely real or imaginary part, and
we also apply the $3^{rd}$ order polynomial filter given in Ref. \cite{YNIB06} to reduce noise when
computing the Fourier transform.  We observe two peaks, representing
each of the two states in the second multiplet ((0,2),(2,0)). The FG
therefore does capture states of double-excitation character.  The
exact values of the (0,2) and (2,0) excitations are indicated on the
graph as dashed lines. We see that FG gives the peak at $\omega=2$
a.u. exactly right and the other is shifted lower to approximately
$\omega=1.58$ a.u.. The peak at $2$ a.u.  is purely an excitation in
the center of mass coordinate, where, as mentioned, the system is
purely harmonic. For harmonic systems, most semiclassical methods,
including FG, are exact~\cite{H81,V28}, and so it is expected
that FG works well for this peak. 
%We also see peaks in the FG
%corresponding to higher-order response which are also surprisingly
%good. {\bf(*peter, please clarify which ones*) - drop this?}

 We reiterate that TDDFT in its usual adiabatic approximation, only
 yields {\it one} peak; Ref. \cite{EGCM11}, examined this pair of
 excitations using adiabatic exact exchange within the single-pole
 approximation and found, as expected, a single peak (of frequency 1.87 a.u.).

A quadratic perturbation yields zero transition probability to states of odd parity, such as in the third multiplet. To access these states, we cubicly {\it kick} the system and
find the third-order moment, also plotted in  Fig.~\ref{f:hookqspec} (dashed line).   This is a harder test for FG as we are
amplifying any errors far from the center in the decaying part of the wavefunction. Again we see that FG works well, giving three peaks corresponding to the first excitation, and the mixed single- and double-excitations in the third multiplet, with again the center-of-mass excitations  given exactly. Looking at
Table \ref{t:hooke}, we see that the FG correctly reproduces the property that the pair of peaks in the third multiplet differs from those in
the second by a single excitation in the center-of-mass
coordinate.

The peaks in the fourth multiplet are most clearly resolved by
applying a quartic kick to the system. These three states involve
mixtures of a single-excitation with two double-excitations, and FG
provides a good approximation. 
%numerical issues where even
%moments are often more sensitive to errors. The $(4,0)$ peak appears
%clearer if the initial monte-carlo sampling is changed slightly, and
%using more trajectories overall will reduces this error.
 
Our spectra could be made cleaner by running for longer times or using
harmonic inversion methods for example. However they are adequate for
our current purposes of illustrating that the FG method captures
correlated states of mixed single and double-excitation character.

In conclusion, although the original motivation for a semiclassical
description of electron correlation was to address challenges TDDFT
has for real-time dynamics in non-perturbative fields~\cite{RRM10},
rather than for spectra, this study shows that semiclassical dynamics
may nevertheless also be useful in the linear response regime. In
particular, FG dynamics does capture states of double-excitation
character, albeit approximately, missing in the usual adiabatic approximation of TDDFT. In
comparison to the ``dressed TDDFT'' that uses a frequency-dependent
kernel derived in Ref.~\cite{MZCB04}, the FG results are not as
accurate, e.g. for the ((0,2),(2,0)) multiplet, dressed TDDFT gives 1.712 a.u. and 2.000 a.u.
On the other hand, in the dressed TDDFT approach, the
procedure to obtain the double-excitations involves an identification
by the user of zeroth-order single- and double-excitations that are
likely to interact strongly, while in the present semiclassical
approach, they emerge naturally from the dynamics. 

We make two remarks at this point. First, a full HHKK treatment would
likely yield more accurate semiclassical spectra, given that the
prefactor missing in FG is complex, so its phase contributes
interference effects important in determining the resonant frequencies.  Second it
is important to note that the proposed method of Ref.~\cite{RRM10}
treats {\it only} the correlation component of the density-matrix
dynamics semiclassically, rather than of the entire dynamics as we
have done here; this may result in improved accuracy of the spectra and
we shall investigate this in the future. Given the more favorable scaling
with system-size of FG than of HHKK, and given that it will ultimately
be used in conjunction with exact kinetic, Hartree, and exchange
components of the evolution, it is the FG method we are most interested in at present.

\subsection{Hooke Model: Natural Orbitals and Momentum Distributions}
\label{sec:HookeNOpdist}
Next, we apply a resonant driving perturbation to the Hooke model dot, to investigate whether FG yields time-dependent NO occupation numbers accurately. Due to the harmonic potential theorem~\cite{D94}, applying an electric field does not change the natural orbital occupation numbers; instead
we apply a quadratic perturbation, $\delta v\ext(x,t) = k(t)x^2/2$, with a time-dependent spring constant:
\ben
%k(t) = 1 - 0.05\sin(2 \omega t)
k(t) = 0.05\sin(2 \ t)
\label{hookedrv}
\een
designed to encourage population transfer to the $(2,0)$ state.
% which is the first accessible due to symmetry. 

In Ref. \cite{RRM10}, the NO occupation numbers for the Moshinsky model were
calculated, however the Moshinsky Hamiltonian  is completely
quadratic and so, as mentioned, FG is exact. In the Hooke model
case, the soft-Coulomb interaction between the electrons makes it a little
more realistic model of a quantum dot, and a harder test for FG. In the following FG calculations for the system being driven with Eq. (\ref{hookedrv}) , only $60000$ classical trajectories were needed for convergence.

\begin{figure}[t]
    \includegraphics[width=7cm,clip]{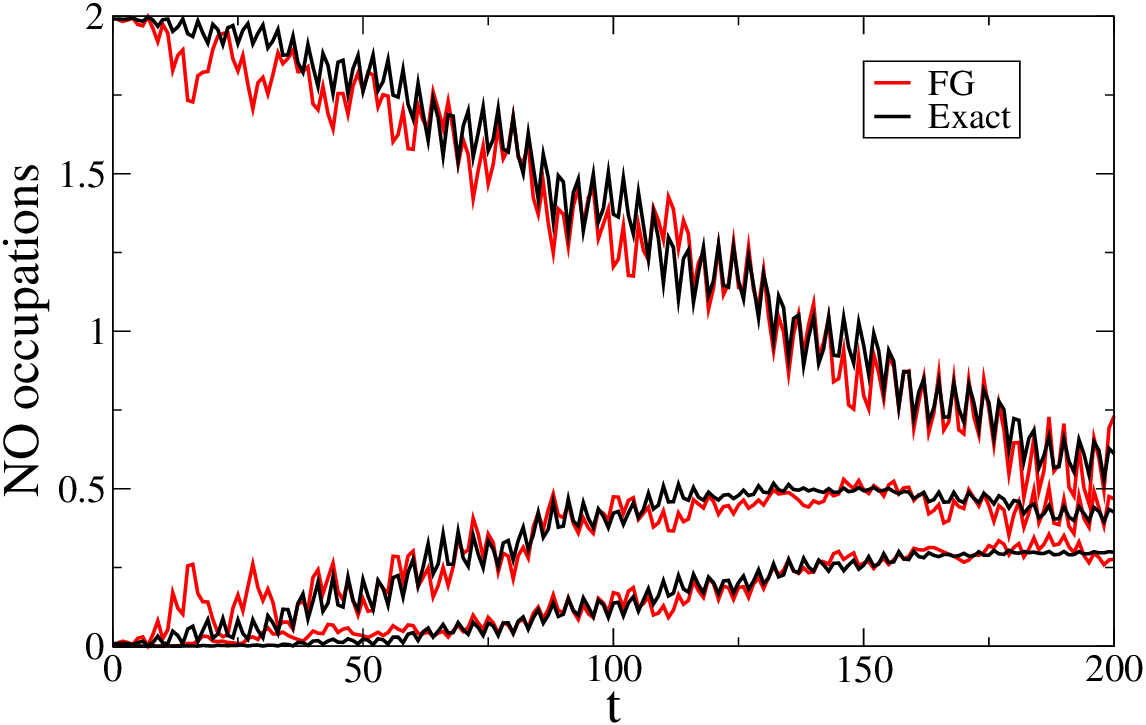}
  \caption{\label{f:hookNO}NO occupation numbers of the Hooke model dot driven at a resonant frequency of $2.0$a.u.}
\end{figure}

\begin{figure}[t]
    \includegraphics[width=8cm,height=4cm,clip]{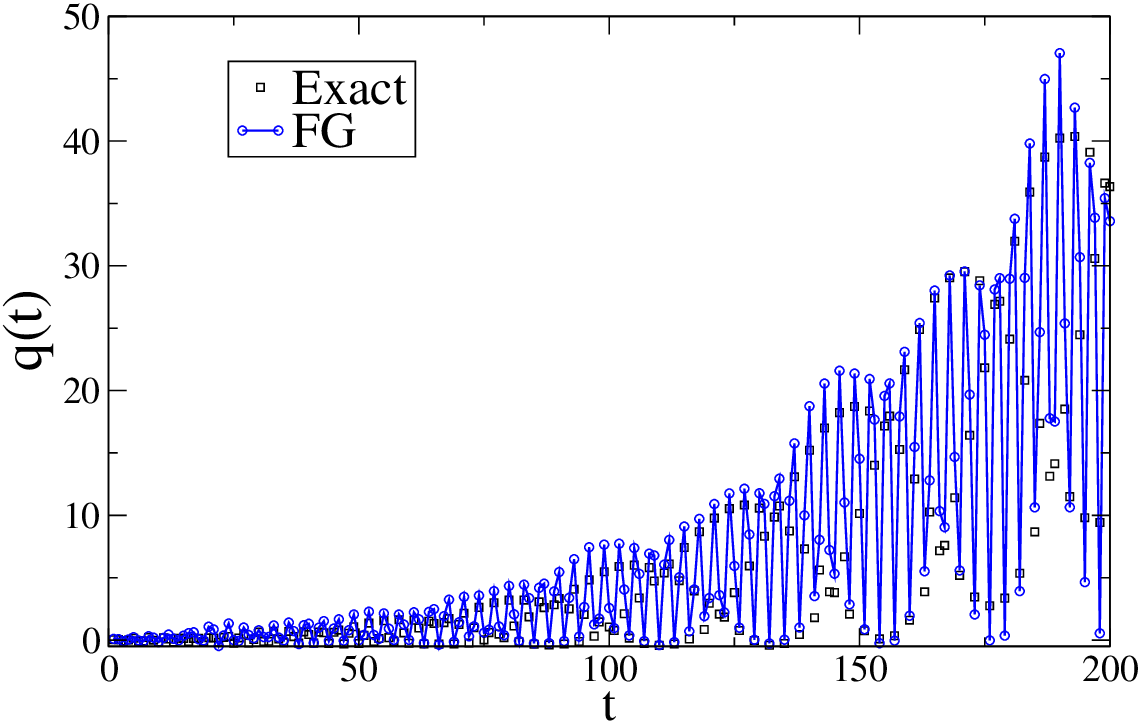}
  \caption{\label{f:hook_quad}The quadrupole moment for Hooke model dot driven at resonant frequency. Due to its highly oscillatory behavior, we show the exact values only as square points for clarity.}
\end{figure}

%\begin{figure}[tb]
%\unitlength1cm
%\begin{picture}(12,6.2)
%\put(-6.2,-3.6){\makebox(12,6.2){
%\special{psfile=hook_fg_cmpNOocc.eps hoffset=0 hscale=35 vscale=35}
%}}
%\end{picture}
%\caption{NO occupation numbers of the Hooke model dot driven at a resonant frequency of $2.0$a.u.}
%\label{f:hookNO}
%\end{figure}
In Fig. \ref{f:hookNO}, we plot the exact and FG NO occupation numbers. Recall that for both exact TDDFT and all
common approximations in TDDMFT, the occupation numbers are fixed by
their initial values and cannot change. It is quite clear that despite
not being perfect, FG works very well here, tracking the large changes in the occupation values.
The quadrupole moment is shown in  Fig. \ref{f:hook_quad}, showing that the FG density is also very accurate. 

%\begin{figure}[tb]
%\unitlength1cm
%\begin{picture}(12,7.2)
%\put(-6.2,-3.6){\makebox(12,7.2){
%\special{psfile=hook_cmp_quad_EX_FG_T200.eps hoffset=0 hscale=45 vscale=45}
%}}
%\end{picture}
%\caption{The quadrupole moment for Hooke model dot driven at resonant frequency. Due to its highly oscillatory behavior, we show the exact values only as square points for clarity.}
%\label{f:hook_quad}
%\end{figure}

In an exact TDDFT KS calculation, the density would be exact, but the
NO occupation numbers would remain at integers 2 (occupied orbital),
and 0 for all the rest. We now show how this feature drastically affects observables that are not directly related to the coordinate-space density. We shall focus on the one-body momentum distribution, defined as 
\ben
n(p;t) = \frac{2}{(2\pi)^2} \int dp' \left\vert \int dxdx' e^{i(px+p'x')} \Psi(x',x;t) \right\vert^2\,,
\een
i.e. it is the probability of finding any one electron with momentum
$p$. Although it is known that the momentum distribution of the exact
KS wavefunction is not that of the true system~\cite{WB07,RHGM09},
little is known about how to extract the latter from the former. 
In the absence of a good observable-functional for momentum, one resorts to simply using the exact KS momentum distribution.

\begin{figure}[tb]
\unitlength1cm
\begin{picture}(12,7.2)
\put(-6.2,-3.6){\makebox(12,7.2){
\includegraphics{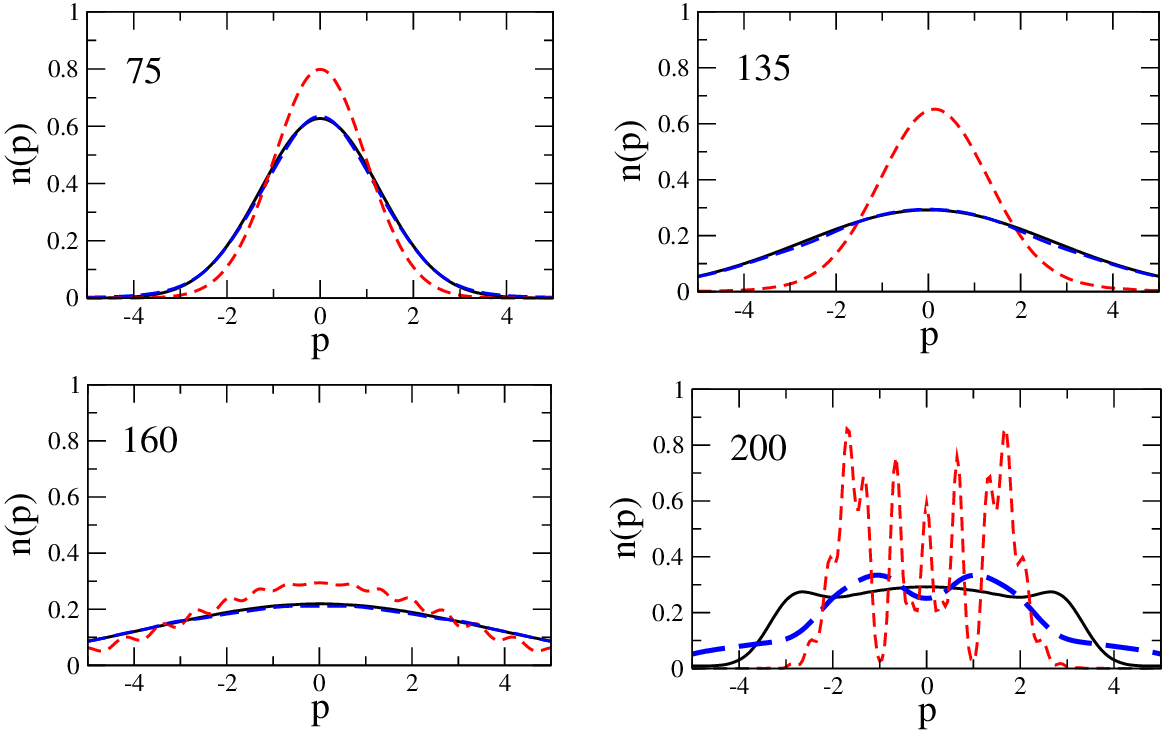}
}}
\end{picture}
\caption{The exact (solid line), KS (short-dashed line), and FG (long-dashed line) momentum densities calculated at times $75$, $135$, $160$, and $200$ au for the on-resonant hooke system. }
\label{f:hooknp200}
\end{figure}

\begin{figure}[t]
    \includegraphics[width=8cm,height=4cm]{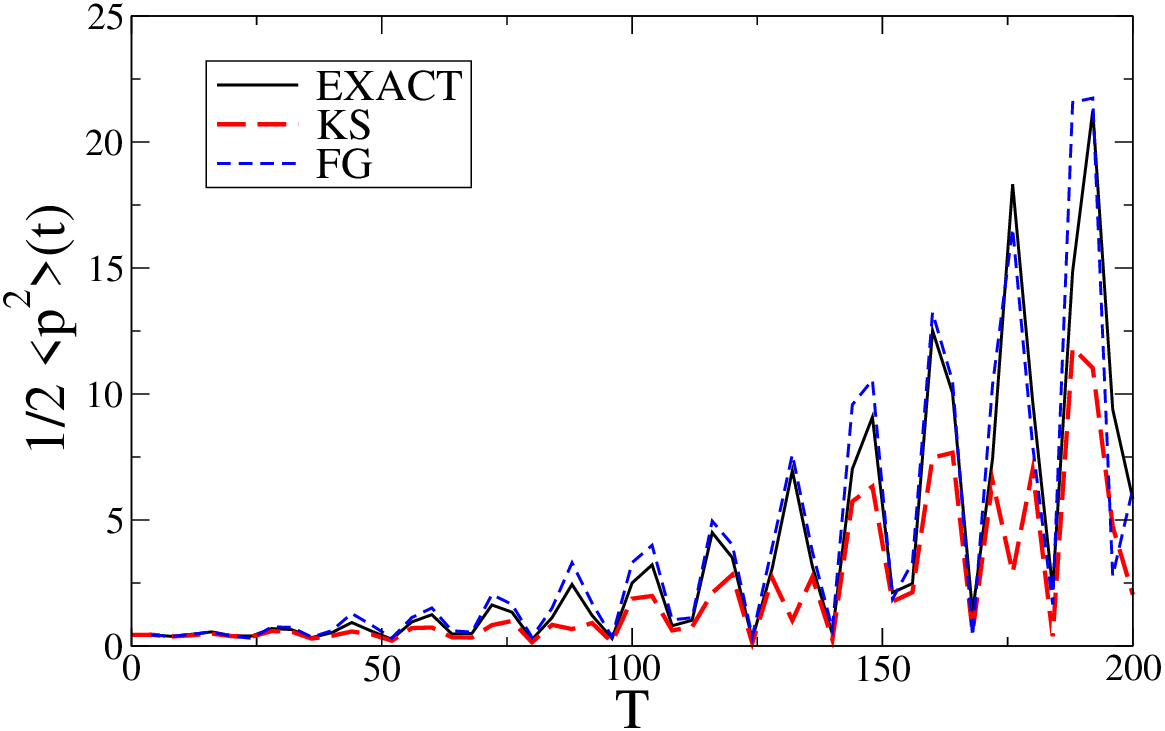}
  \caption{\label{f:hookKE}The kinetic energy  for the exact, KS, and FG calculations, when the Hooke dot is driven on resonance. Although the KS system must reproduce the exact momentum, the kinetic energy becomes worse for longer times, while the error in the FG remains about the same.}
\end{figure}

In Fig. \ref{f:hooknp200}, we show the exact, exact KS
(computed directly from the exact KS orbital obtained via the
procedure in Sec \ref{sec:results}), and FG momentum distributions at four snapshots in time.
The first moment of the KS and exact distributions
must integrate to the same value for all one-dimensional systems (in
this case zero), due to the facts that 
$\langle p(t) \rangle = \int dx j(x,t) = \int dx x \dot{n}(x,t)$, 
%$\langle p(t) \rangle = \int dx j(x,t) = \int dx \int^x dx' \dot{n}(x',t)$, 
and that the KS exactly tracks the density.  Despite
this, the KS distribution oscillates wildly. The reason for
this is not dissimilar to that seen in Ref.~\cite{RHGM09}: the
KS system becomes strongly non-classical as it attempts to
describe the system of two-electrons using a single (doubly-occupied)
orbital. For short times, the KS $n(p)$ behaves well, however as the
breathing motion of the system becomes more pronounced, moving out
further and faster, it becomes an increasingly non-classical dynamics
for a single orbital, such as in KS, to describe. The
underlying phase-space distribution of the KS system develops
strong oscillations into negative regions, signifying the
non-classical description of different ``parts'' of one electron
moving in different directions.  
The momentum-distribution represents one observable that the occupation numbers strongly influence. The FG results are much more accurate than the exact KS, as is also reflected in the kinetic energies shown in Fig. \ref{f:hookKE}. 

%\begin{figure}[tb]
%\unitlength1cm
%\begin{picture}(12,7.2)
%\put(-6.2,-3.6){\makebox(12,7.2){
%\special{psfile=hook_cmp_p2mom_EX_KS_FG.eps hoffset=0 hscale=35 vscale=35}}}
%\end{picture}
%\caption{The kinetic energy  for the exact, KS, and FG calculations, when the Hook dot is driven on resonance. Although the KS system must reproduce the exact momentum, the kinetic energy becomes worse for longer times, while the error in the FG remains about the same.}
%\label{f:hookKE}
%\end{figure}

\subsection{Soft-Coulomb Helium: The Problem of Classical Autoionization}
\label{sec:sftcAI}
Now we move from model quantum dots to model atoms: a soft-Coulomb interaction is used for the nuclear potential, as a one-dimensional model of a Helium atom (see e.g. Refs.~\cite{JES88,LL98,BN02,LGE02,LK05}). The Hamiltonian is given by:
\bea
\nonumber
\hat{H} &=&\sum_{i=1}^2\left(-\half\frac{d^2}{dx_i^2}- \frac{2}{\sqrt{x_i^2+1}} + \epsilon(t)x_i\right) \\
&+& \frac{1}{\sqrt{(x_1-x_2)^2+1}} 
\label{eq:Hsftc}
\eea
where the third term on the first line represents an external electric
field applied to the system. This model has been used in several
studies of correlated electron dynamics in strong fields as well as in
TDDFT, since it allows a numerically exact solution to be calculated
relatively easily.

\begin{figure}[t]
    \includegraphics[width=6cm,height=6cm,clip]{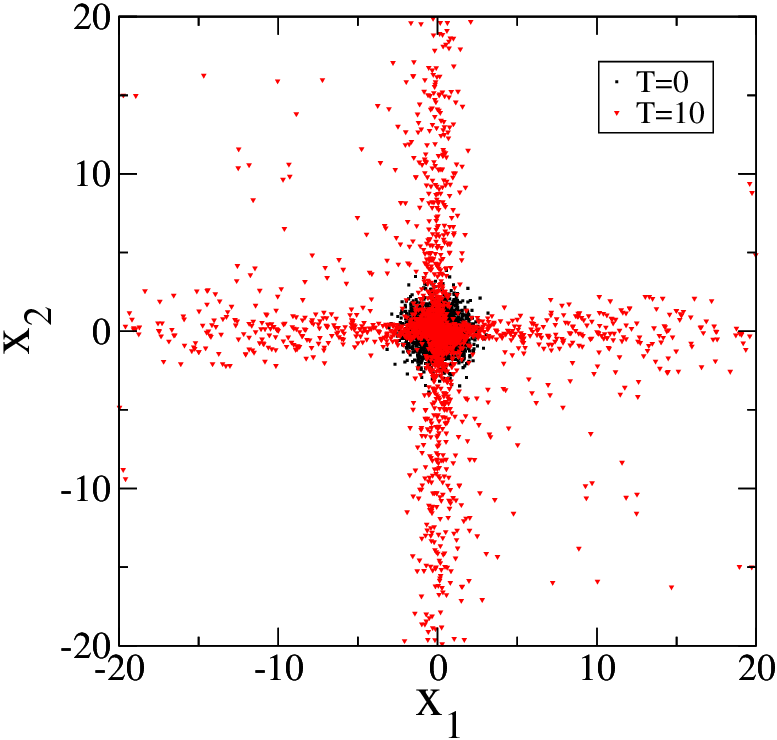}
  \caption{\label{f:sftcnofieldtraj}The positions, $x_1$ and $x_2$, of both particles plotted against eachother at the initial time (square points) and at time $t=10$ a.u. (triangle points). The problem of classical autoionization is characterized by the cross-shape distribution at the later time: one particle falls to the center while the other flies away from the atom.}
\end{figure}

\begin{figure}[t]
    \includegraphics[width=8.5cm,height=6cm]{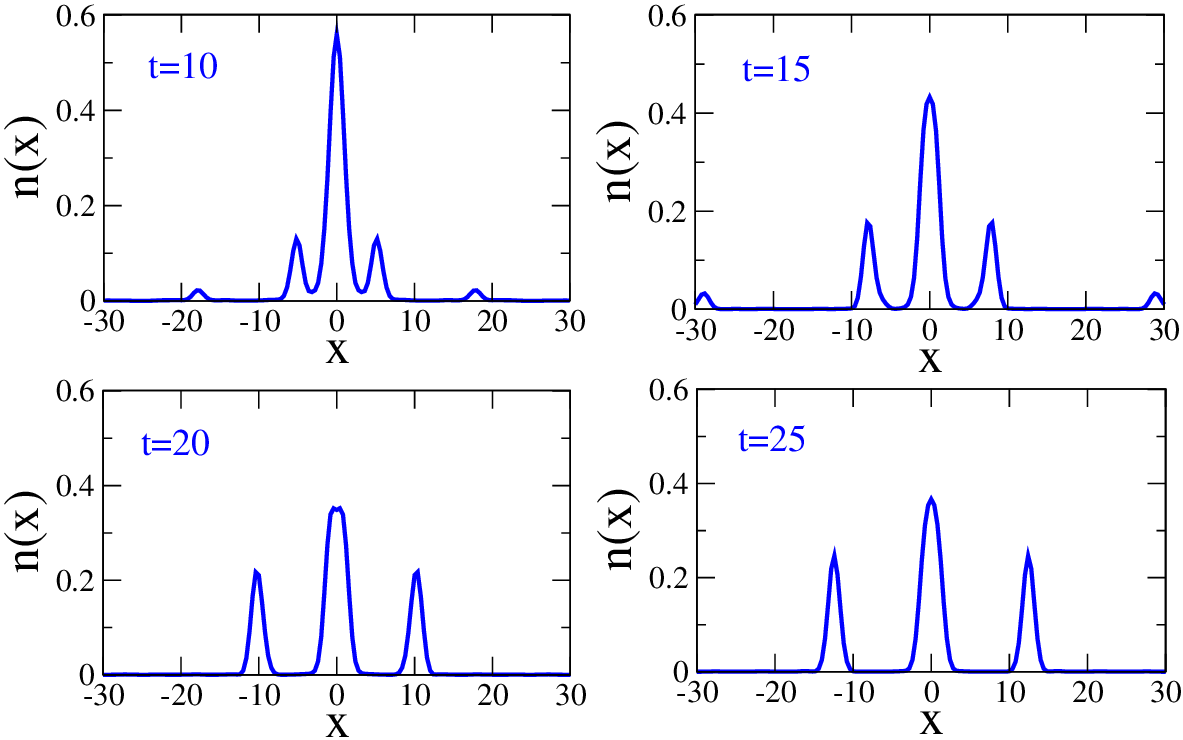}
  \caption{\label{f:sftcFGden_unscr}Snapshots of the FG density at various times when no field is applied to the soft-Coulomb He atom. The lobes on either side of the central distribution represent autoionizing wavepackets. The number of trajectories in the monte carlo sum was $M=20000$.}
\end{figure}

\begin{figure}[t]
    \includegraphics[width=6cm,height=4cm]{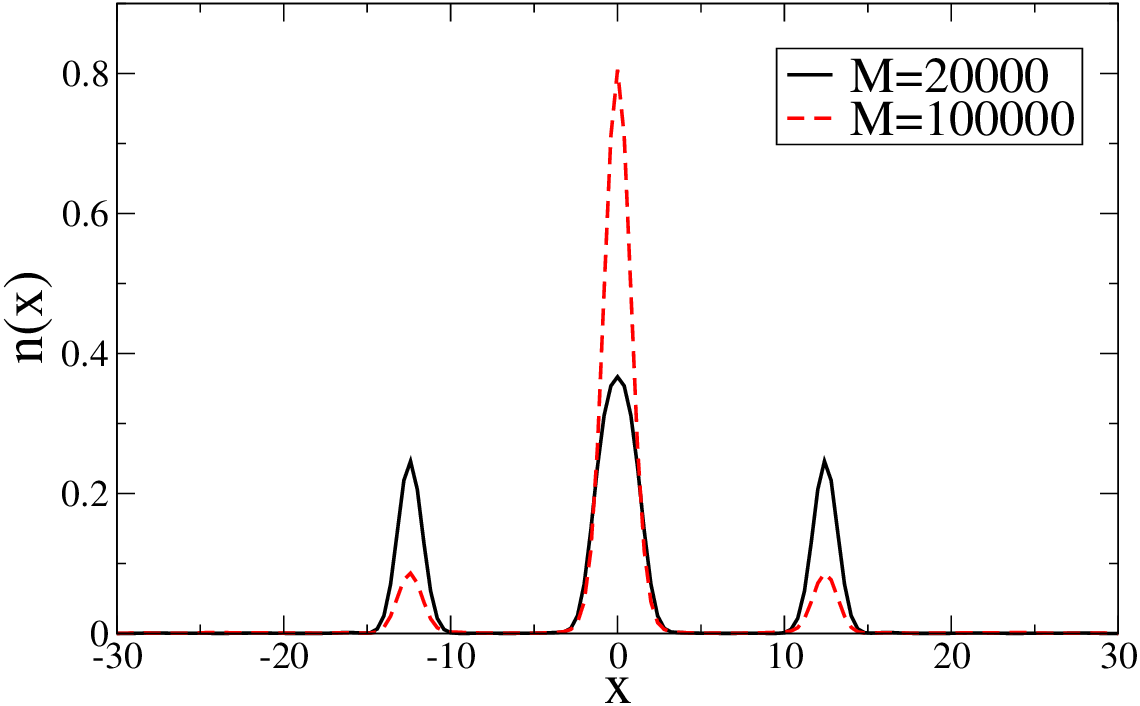}
  \caption{\label{f:sftc_NOFIELD_FG_cmpden}The FG density at time $t=25$ a.u. (with no field applied to the soft-Coulomb He atom). Adding more trajectories reduces the classical autoionization.}
\end{figure}

With our electronic system no longer everywhere bound, we 
encounter the problem of classical autoionization, mentioned in the
introduction. Even when {\it no} field is present, this haunts the
dynamics, as we will now show.  We begin in the exact ground-state of
the Hamiltonian Eq.~(\ref{eq:Hsftc}) with $\epsilon(t) =
0$. Fig. \ref{f:sftcnofieldtraj} plots the positions $x_1,x_2$ of the
trajectories in the initial distribution, and where they have evolved
to at $T=10$a.u. The initial distribution centered at the origin
increasingly evolves into a cross, signifying the classical
autoionization: one electron zipping off to infinity after stealing
the energy from the other left near the origin.  Using just 20000
trajectories to compute the FG sum, we plot in
Fig.~\ref{f:sftcFGden_unscr} the density at four snapshots in
time. Autoionizing wavepackets are clearly seen, evolving out
quasiclassically on either side of the central distribution. (Note
that in the exact problem, nothing happens; the distribution remains
centered at the origin, as the initial state is an eigenstate.)
Although it may appear that the autoionizing packets are growing as a
function of time, this is an artifact of the way the FG dynamics is
renormalized (see Sec.~\ref{sec:BackgroundFG}). Before renormalization, the FG norm decreases with time as more trajectories from the central distribution are lost to autoionization and other incoherent effects. 
Why some of the classically autoionizing
trajectories seem to add coherently, forming the wavepackets on the sides, remains to be understood. However
choosing a different initial seed to generate the initial distribution
results in autoionizing wavepackets centered at different positions moving with different speeds. That is,
certainly the results in Fig.~\ref{f:sftcFGden_unscr} are
  unconverged. In fact, in Fig.~\ref{f:sftc_NOFIELD_FG_cmpden} we show
  the effect of increasing the number of trajectories is to
  significantly decrease the classical autoionization. We deduce that by phase
  interference in the semiclassical sum, contributions
  from classically autoionizing trajectories cancel each other out.
Considering that HHKK-semiclassics gives the
  correct dynamics to order $\hbar$, this is as it should be: Although FG dynamics is not
  correct to order $\hbar$ a similar effect happens here.

Although these results show that classical autoionization is not a
fundamental problem of the semiclassical method, but rather a
convergence issue, it does remain an important practical one for most
applications; impossibly large numbers of trajectories could be
required to make the classical autoionization effect small enough for
the purposes. In our following studies we terminate trajectories in
one of two ways. In the applications we consider, we do not expect
significant amplitude beyond a certain distance. So, (i) we simply
terminate any classical trajectory where the coordinate of one
electron has reached a prescribed distance, $L$, away from the
nucleus.  This is similar to what is done in Ref.~\cite{HK07} in
obtaining semiclassical spectra (where the termination distance was
additionally dependent on the total energy of the trajectory). Another alternative is to use purely the energy terms to determine autoionization such as was done in Ref.~\cite{RTW93}. 
At all times, this means that we may still observe
autoionizing wavepackets peeling off from the central distribution,
that eventually reach $L$, only after which they are discarded;
therefore at the earlier times, these may erroneously contribute to
the FG. We therefore also consider (ii) prescreening all trajectories,
such that if they, at some point during the duration of interest,
reach $L$, they are discarded from the very start of the
propagation. (Note this prescreening process is very fast, as only classical dynamics needs to be run; therefore many  initial candidate trajectories may be explored). In this way no autoionizing trajectories (as defined by
$L$) ever contribute to the FG sum.  In our particular studies, there
was some difference between the results obtained with (i) and (ii)
especially at shorter times, with (ii) leading to narrower
distributions than (i). It is unclear to us which way is more
``correct'', especially given the fact that, theoretically, none of
the trajectories are ``wrong'': including all, together with their
phases, there is no classical autoionization problem. In particular,
does the earlier behavior of a trajectory that eventually autoionizes,
contribute in a non-cancelling, sensible way to the dynamics at times prior to its autoionizing event? This is an important
question we leave for future work.

%{\bf peter - you had  ``GRAPH: compare prescreened to others (waiting for this to finish)'', are we still waiting on this? Compared a nofield 2mil prescrn to a renormed unscrn, not impressive enough imo}

%\begin{figure}[tb]
%\unitlength1cm
%\begin{picture}(12,6.2)
%\put(-6.2,-3.6){\makebox(12,6.2){
%\special{psfile=sftc_NOFIELD_FG_cmp_den_t25.eps hoffset=0 hscale=40 vscale=40}
%}}
%\end{picture}
%\caption{The FG density at time $t=25$ a.u. (with no field applied to the soft-Coulomb He atom). Adding more trajectories reduces the classical autoionization.}
%%\label{f:sftc_NOFIELD_FG_cmpden}
%\end{figure}

%\begin{figure}[tb]
%\unitlength1cm
%\begin{picture}(12,6.2)
%\put(-8.2,-5.9){\makebox(12,6.2){
%\special{psfile=sftc_NOFIELD_FG_M20000_wig_t25.eps hoffset=0 hscale=85 vscale=85}
%}}
%\end{picture}
%\caption{unscrn nofield FG wigner function for $M=20000$ at time $t=25$ a.u.}
%\label{f:sftc_NOFIELD_wig_M20000}
%\end{figure}

%\begin{figure}[tb]
%\unitlength1cm
%\begin{picture}(12,6.2)
%\put(-8.2,-5.9){\makebox(12,6.2){
%\special{psfile=sftc_NOFIELD_FG_M2mil_wig_t25.eps hoffset=0 hscale=85 vscale=85}
%}}
%\end{picture}
%\caption{unscrn nofield FG wigner function for $M=2000000$ at time $t=25$ a.u.}
%\label{f:sftc_NOFIELD_wig_M2mil}
%\end{figure}

\subsection{Soft-Coulomb Helium:Strong field}
\label{sec:sftctrap}
\begin{figure}[t]
    \includegraphics[width=7cm,height=4cm]{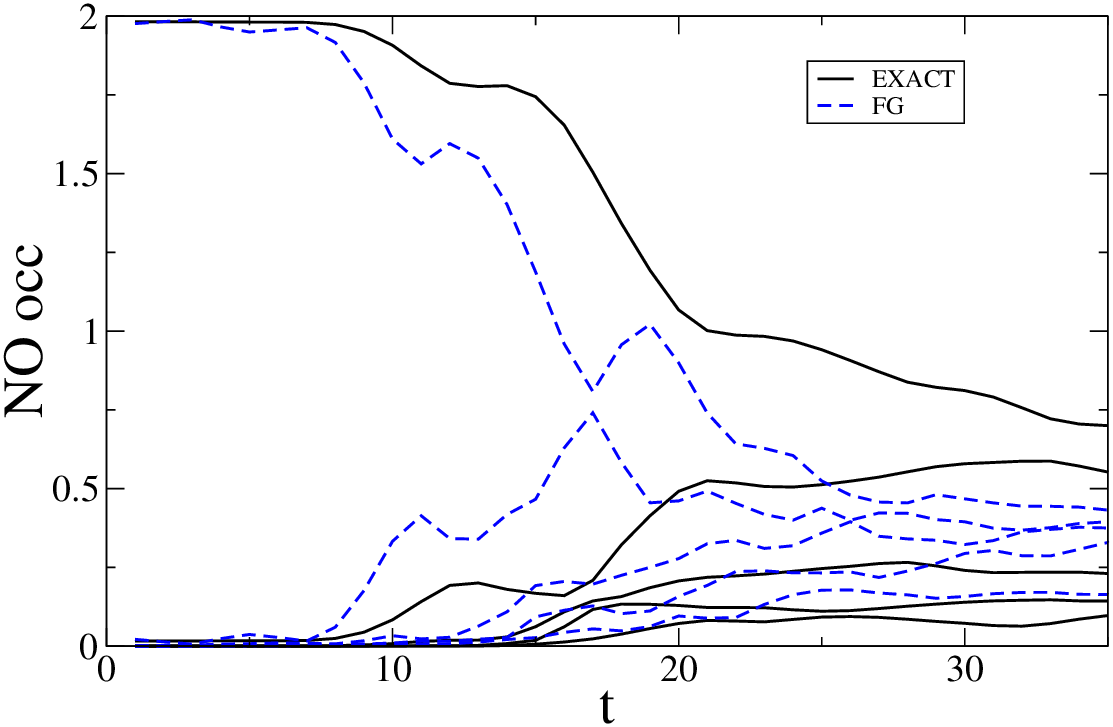}
  \caption{\label{f:sftcEtrapNO}NO occupation numbers for the soft-Coulomb He atom dynamics induced by Eq.~\ref{eq:trapeps}.}
\end{figure}

\begin{figure}[t]
    \includegraphics[width=7cm,height=4cm]{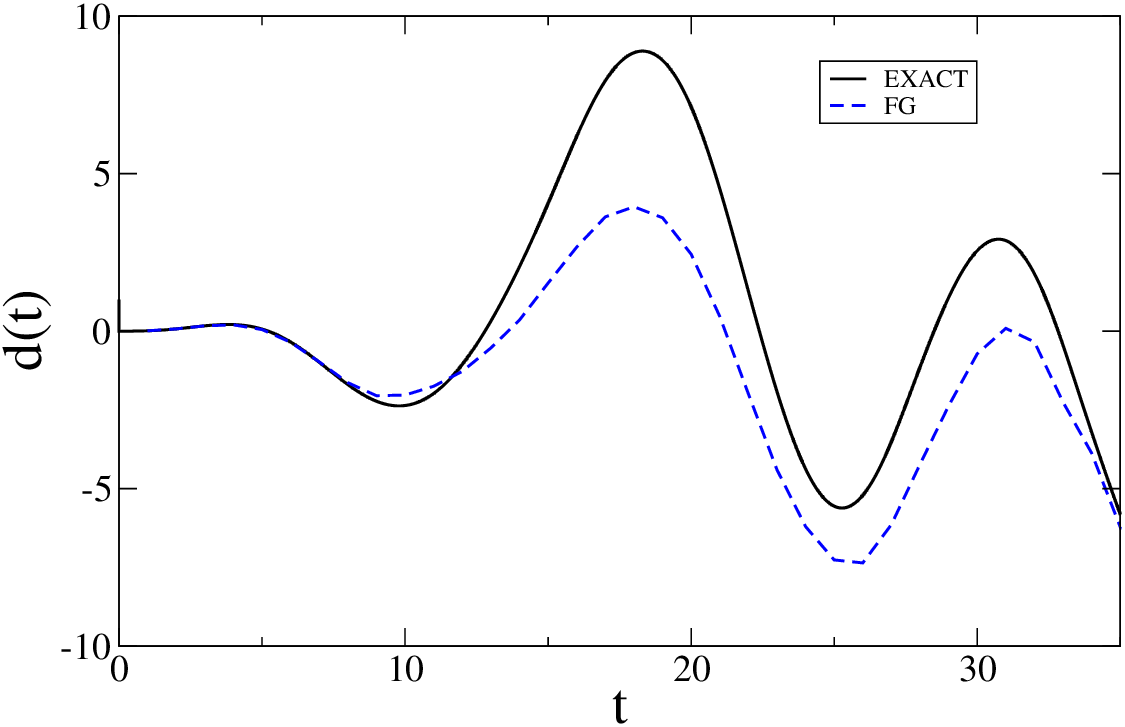}
  \caption{\label{f:sftcEtrapdip}Comparing the FG dipole moment in the soft-Colulomb He dynamics under Eq.~\ref{eq:trapeps}.}
\end{figure}

\begin{figure}[t]
    \includegraphics[width=8.5cm,height=7.5cm]{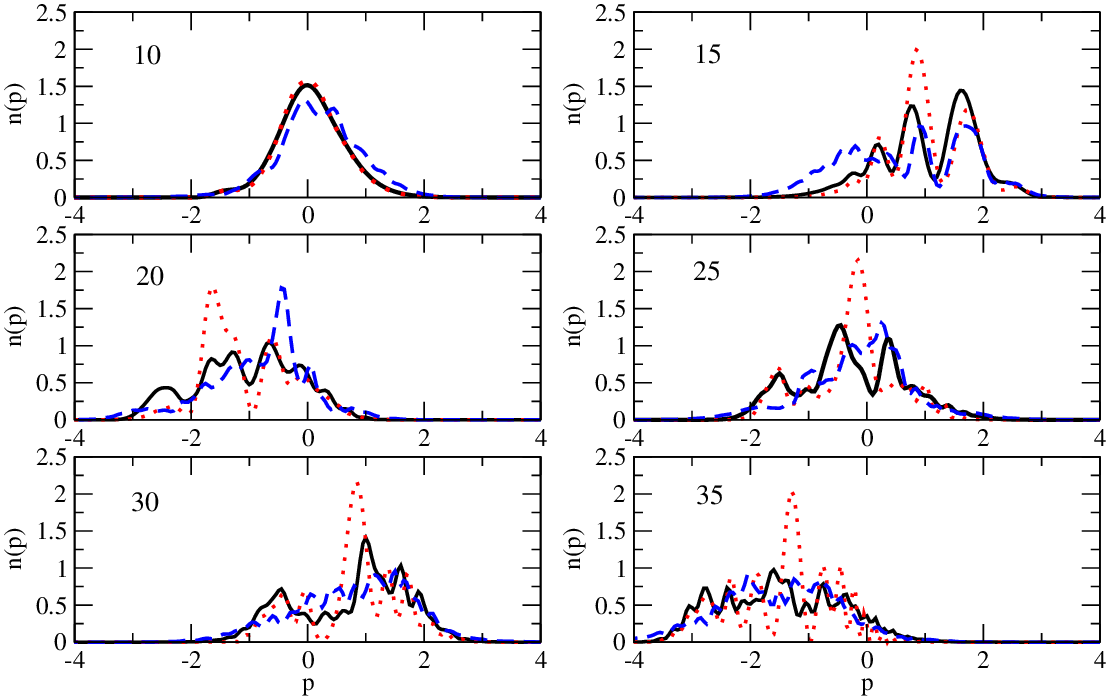}
  \caption{\label{f:sftcnp}The momentum density n(p) at times $10,15,20,25,30$ for the strong field case. Solid line is exact, dotted is KS, and short-dashed is FG.}
\end{figure}

We now apply a relatively strong field to the model He atom:
\ben
\epsilon(t) = \frac{1}{\sqrt{2}}\cos(0.5t) \left\{\begin{array}{c l} \frac{t}{20} & t\leq 20 \\ 1 & t>20 \end{array}\right.
\label{eq:trapeps}
\een 
where the electric field is ramped-up linearly until $t=20$ au. The
value of this field is large enough that classical autoionization
effects are relatively small in comparison to the large oscillations
in the dynamics induced by the field.  
We use the procedure (i) described above and terminate trajectories once $\vert x_i\vert > 30$a.u. We found the results were essentially converged starting with $2\times 10^6$ trajectories and finishing with $5\times 10^5$, the rest being discarded by the termination condition at some point during the run.
%{\bf also, can we comment on how results compare with unscreened, esp at shorter times?}
We plot the NO occupation
numbers in Fig.~\ref{f:sftcEtrapNO}, up to a time of 35 a.u. Notice
that the number of trajectories needed for convergence in the
soft-Coulomb He atom is much larger than those needed in the
Hooke-model quantum dot; the latter is almost a best case scenario for
semiclassics, because of the quadratic nature of the external
potential. The FG captures the change in occupation numbers, perhaps a little
too enthusiastically; Figure~\ref{f:sftcEtrapdip} shows the dipole is
also reasonably well approximated. In Fig.~\ref{f:sftcnp}, we plot the
momentum distributions at several snapshots of time, comparing the FG
with the exact, and with the exact-KS, as in
Sec.~\ref{sec:HookeNOpdist}. We see that the error of the FG remains about
the same throughout the evolution, while the error of the KS increases
with time. This is a direct consequence of the change in NO occupation
numbers: the KS momentum distribution is that of a SSD with NO
occupation numbers 2 for one orbital and zero for all others, while
that of the true and, roughly captured by FG, changes dramatically in
time, as shown in Fig.~\ref{f:sftcEtrapNO}. Consequently, typical observables not directly related to the density, when evaluated using the usual operators on the KS wavefunction become badly approximated; the unknown observable-functionals of exact TDDFT are likely quite complicated.

%\begin{figure}[tb]
%\unitlength1cm
%\begin{picture}(12,5.9)
%\put(-6.2,-3.45){\makebox(12,5.9){
%\special{psfile=sftc_cmp_np_timeslice_EX_KS_FG_6times.eps hoffset=0 hscale=45 vscale=45}
%}}
%\end{picture}
%\caption{The momentum density n(p) at times $10,15,20,25,30$ for the strong field case. Solid line is exact, dotted is KS, and short-dashed is FG.}
%\label{f:sftcnp}
%\end{figure}

\subsection{Soft-Coulomb Helium: Towards Optimal Control Theory}
\label{sec:sftcopt}
\begin{figure}[t]
    \includegraphics[width=7cm,height=4cm,clip]{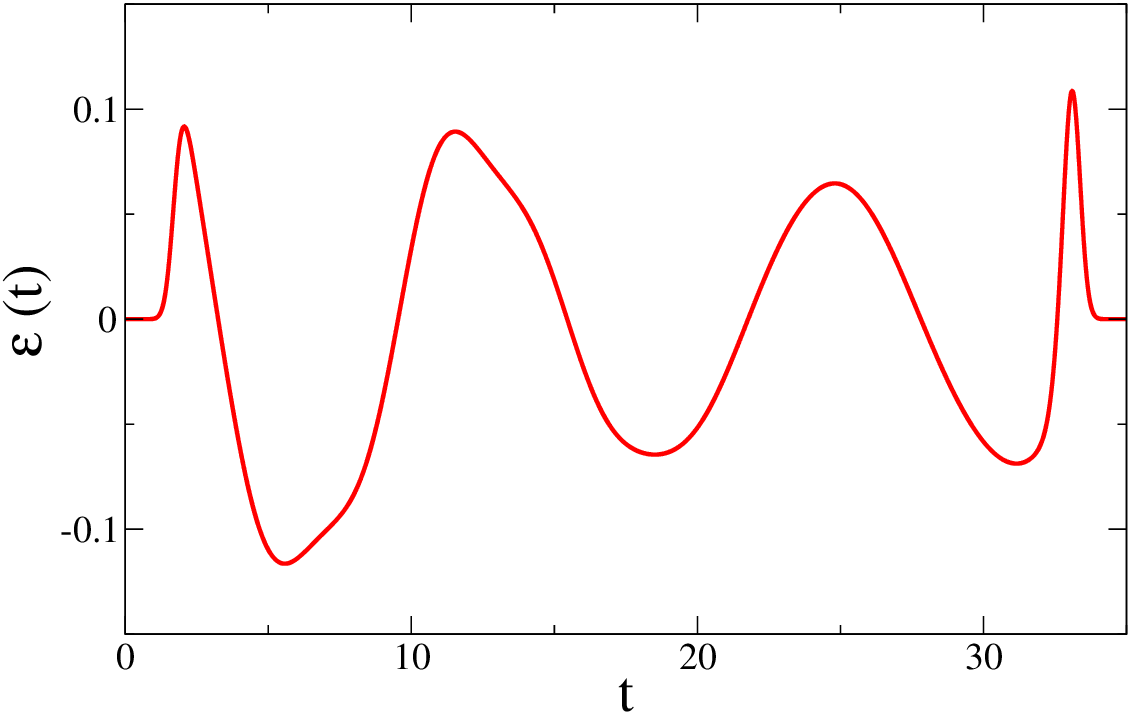}
  \caption{\label{fig:optfield}The optimal electric field to maximally populate the first excited singlet-state of soft-coulomb Helium within a time of $35$au.}
\end{figure}

Our final example is also the hardest one: we apply an optimal control
field to the soft-Coulomb He atom, to try to get it to evolve from the
ground-state to the first-excited state. The occupation numbers in the ground-state are close to 2 for the highest, and close to zero for all others, while in the target state, the highest two occupation numbers are both close to 1, having a double Slater determinant character.
(See also Sec.~\ref{sec:Background}, and~\cite{RRM10,MBW02}). 
However, instead of
optimizing the field for the FG evolution, we simply take an optimal
field found for the exact dynamics, and ask how well does it work for
the FG problem. (Of course, the former is what we ideally would do,
however, we leave the coding of optimal control problems for future
work.) We find the optimal field using the optimal control functionality of the octopus
code~\cite{octopus,CWG11}. Moreover, we consider a relatively short
duration for the optimal pulse, $T=35$ a.u., as we will find this
illustrates already the challenges the FG approach has for such
problems. Because there are only a few cycles of the laser field in
this short time, the yield in the exact problem is not very large: the optimal field found in Fig.~\ref{fig:optfield} yields a final state population of $0.73$..
The field is significantly weaker than in the previous section, and we
really do not expect much ionization, classical or real. To avoid the situation of Fig.~\ref{f:sftcFGden_unscr}, we therefore
use method (ii) to terminate the trajectories, prescreening them such
that no trajectories where one coordinate reaches a distance $L=10$ a.u. at any time in the run are included. We find that beginning with $5\times 10^6$ candidate trajectories, the prescreening process discards many such that $2391950$ remain for the FG calculation.

%\begin{figure}[tb]
%\unitlength1cm
%\begin{picture}(12,6.2)
%\put(-6.2,-3.6){\makebox(12,6.2){
%\special{psfile=SFTC_OPT_T35_laser.eps hoffset=0 hscale=45 vscale=45}
%}}
%\end{picture}
%\caption{The optimal electric field to maximally populate the first excited singlet-state of soft-coulomb Helium within a time of $35$au.}
%\label{fig:optfield}
%\end{figure}

\begin{figure}[t]
    \includegraphics[width=7cm,height=4cm]{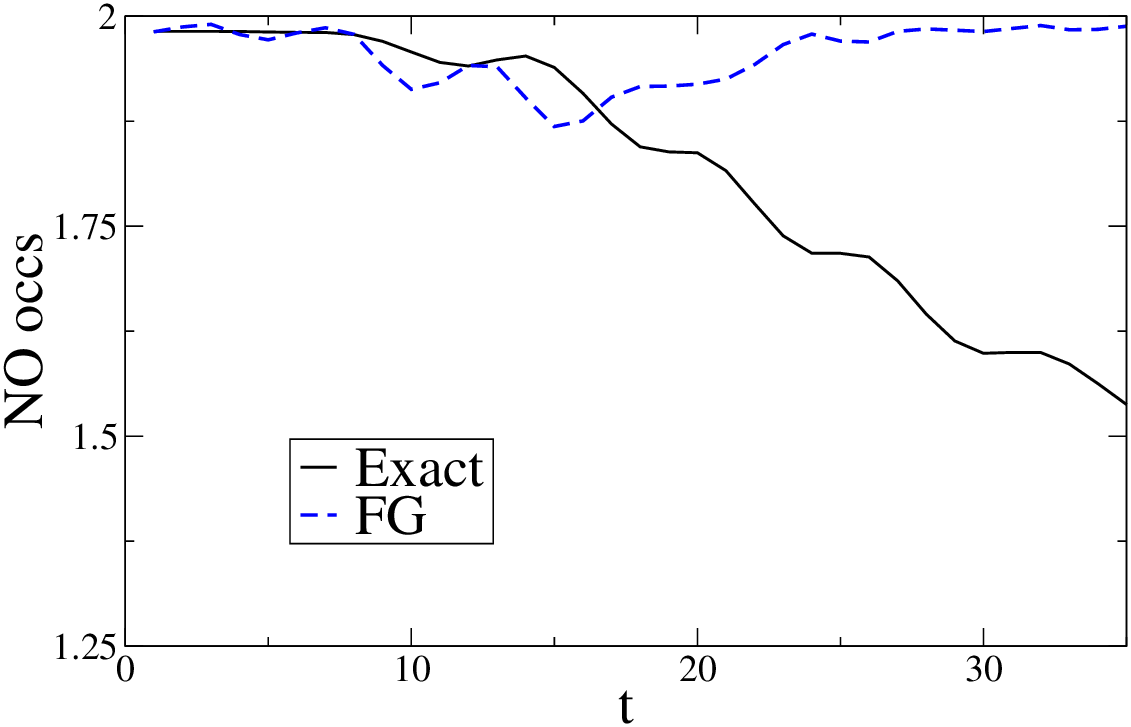}
  \caption{\label{f:sftcoptNO}The highest natural orbital occupation for a prescreened FG calculation with the optimal electric field of Fig. \ref{fig:optfield}. The exact NO value is nearly $1$ at the end, reflecting the double ssd character of the target state.}
\end{figure}

\begin{figure}[tb]
\unitlength1cm
\begin{picture}(12,6.2)
\put(-6.0,-3.6){\makebox(12,6.2){
\includegraphics{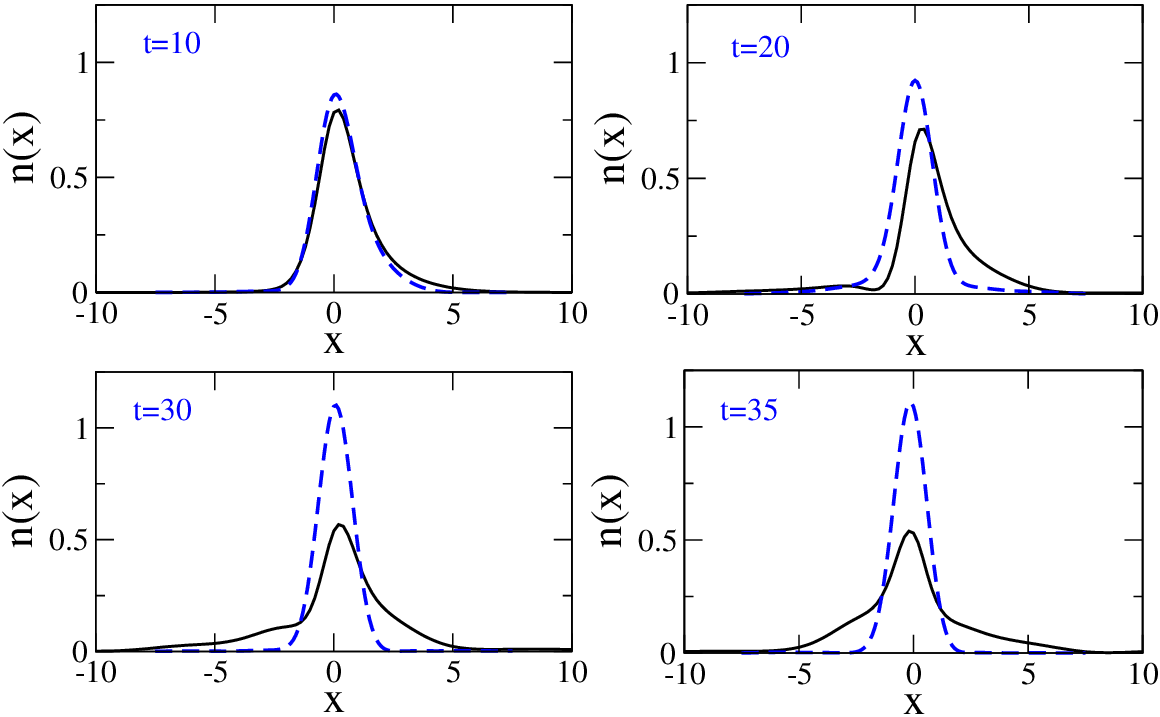}
}}
\end{picture}
\caption{The density of the prescreened FG calculation with the optimal field at various times compared to the exact.}
\label{f:sftcoptdens}
\end{figure}

\begin{figure}[t]
    \includegraphics[width=7cm,height=4cm,clip]{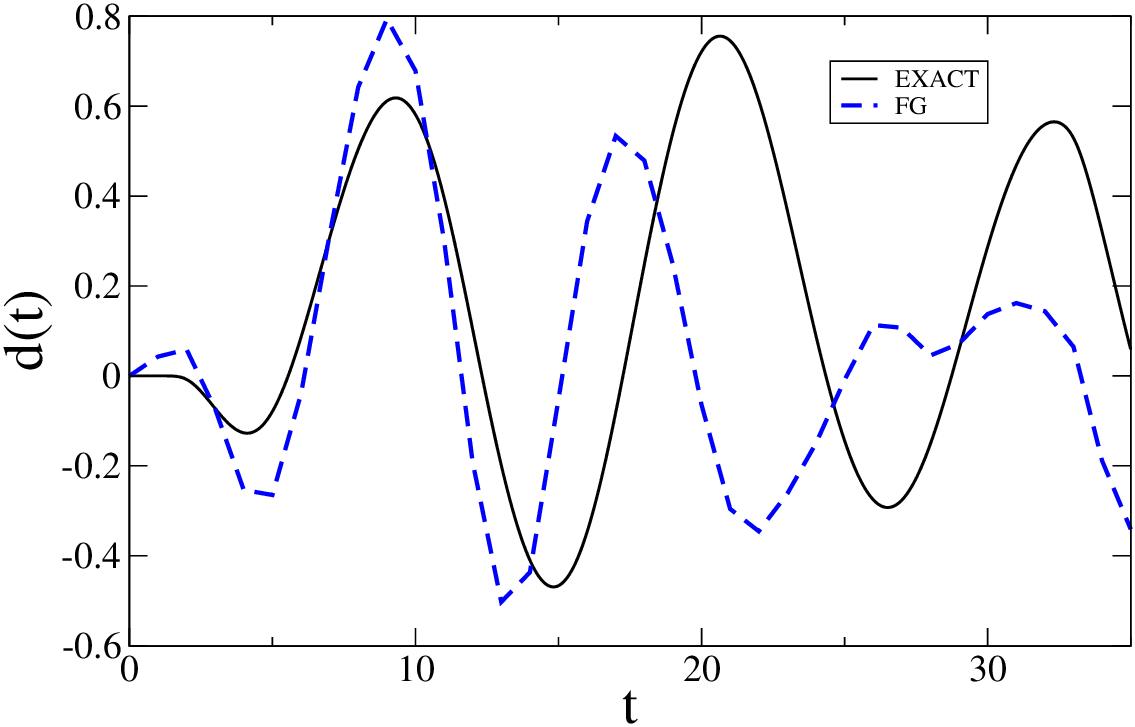}
  \caption{\label{f:sftcoptdip}The prescreened FG dipole moment versus the exact as a function of time. Although initially the FG mimics the true at short times, the delicate nature of quantum control means FG is too inaccurate for this problem.}
\end{figure}

The result for the NO in in Fig.~\ref{f:sftcoptNO} are 
disappointing. The FG results are converged, as different seeds, and
increasing the number of candidate trajectories did not change the
results very much. Also, simply running unscreened calculations with
increasing numbers of trajectories, appeared to converge, more or
less, to these results. Although initially the largest FG NO
occupation number tracks the exact reasonably well, we see that after
about $T=15$, it turns upwards instead of continuing down; the FG
state appears to reduce to a SSD instead of developing more of a double -SD
character. Turning to Figs.~\ref{f:sftcoptdens}
and~\ref{f:sftcoptdip}, we see that after beginning to spread, the
density then contracts and settles into a narrow distribution in the
well. The dipole at shorter times is not bad but fails at larger times.
We also note that the density appears to tighten as time
evolves; this is a consequence of autoionization and occurs even if we
do not apply any termination to the trajectories, in the limit of
convergence; the predominant trajectories that autoionize are those in
the tails of the distribution.

Why the FG results are poor at larger times is, we believe, due to the
FG resonant frequencies being so off-set from the true ones, that a
field that is optimal for population transfer for one is off-resonant
for the other, leading to little transfer. The exact excitation
frequency to the first excited state is $0.53$,
while the FG one is about $0.68$ (found by applying a kick to
the initial ground-state, as in Sec.~\ref{sec:Hookespectra}). 
For example, in the exact case, if the frequency of the applied field is shifted by as little as 0.05au away from resonance, there is no longer the strong change in NO numbers that we saw in Fig.~\ref{f:hookNO}.

Thus the optimal control problem presents an extremely tough test of the frozen gaussian dynamics as it depends critically on subtle interference effects of on-resonance oscillations. As stated in Sec..~\ref{sec:Hookespectra}, we expect the FG excitation frequencies to improve when coupled with the TDDM propagation, and hence also improve the optimal control results. This will be investigated in future work.

%If instead we apply a field of the latter frequency to the FG system, we
%do get a large change in occupation number
%{\bf peter, check? - NO, T100 didnt: except prescreened loads}.

%This appears to be a serious obstacle in the application of FG
%dynamics to optimal control problems, as these problems depend
%significantly on interference effects, quite sensitive to the resonant
%frequencies of the system.

%it is likely that including the complex prefactor of HHKK improves the
%accuracy of the frequencies, but, as also discussed there, we would
%like to avoid this for the many-body problems in mind for the method
%of Ref.~\cite{RRM10}.  Whether or not combining FG for correlation
%{\it alone} with an exact treatment of the one-body terms in
%Eq.~\ref{eq:rho1dot}, as in the prescription of Ref.~\cite{RRM10}, can
%redeem FG to be useful for this type of application depends heavily on
%whether the excitation frequencies of the system improve enough in
%such a combined treatment. This will be investigated in future work.

%\begin{figure}[tb]
%\unitlength1cm
%\begin{picture}(12,6.2)
%%\put(-6.2,-3.6){\makebox(12,6.2){
%\special{psfile=sftc_OPT_T35_cmp_NO_ex_prescrenFG.eps hoffset=0 hscale=45 vscale=45}}}
%\end{picture}
%\caption{The highest natural orbital occupation for a prescreened FG calculation with the optimal electric field of Fig. \ref{fig:optfield}. The exact NO value is nearly $1$ at the end, reflecting the double ssd character of the target state.}
%\label{f:sftcoptNO}
%\end{figure}

%\begin{figure}[tb]
%\unitlength1cm
%\begin{picture}(12,6.2)
%\put(-6.2,-3.6){\makebox(12,6.2){
%\special{psfile=sftc_OPT_T35_cmp_dip_ex_prescrenFG.eps hoffset=0 hscale=45 vscale=45}
%}}
%\end{picture}
%\caption{The prescreened FG dipole moment versus the exact as a function of time. Although initially the FG mimics the true at short times, the delicate nature of the quantum control problem makes it hard for FG to fully describe the situation.}
%\label{f:sftcoptdip}
%\end{figure}

%\begin{figure}[tb]
%\unitlength1cm
%\begin{picture}(12,6.2)
%\put(-6.2,-3.6){\makebox(12,6.2){
%\special{psfile=sftc_OPT_T35_cmp_np_FG_pre_unscrn_t35.eps hoffset=0 hscale=45 vscale=45}
%}}
%\end{picture}
%\caption{OPT comparing $n(p)$ for prescreened and unscreened at $t=35$}
%\label{f:sftcoptnpfg}
%\end{figure}

\section{Conclusions and Outlook}
\label{sec:Conclusions}
The semiclassical FG method provides an intuitive picture
of quantum mechanics, as based on classical trajectories, each smeared out
into a little fuzzy ball in phase-space, evolving classically in time,
and added together with phases determined by their classical
action. It, and its more rigorous HHKK version, has also
provided useful numerical tools in quantum molecular
dynamics. Its application to interacting electronic
systems has been hesitant, although recent work~\cite{HK07} clearly
shows its promise for electronic spectra. The results of the present
work suggest that for systems of interacting electrons in external
fields, FG dynamics can also be useful. 

In relation to TDDFT and adiabatic TDDMFT, we have shown that FG dynamics 
overcomes some of the problems these methods have; capturing states
of double-excitation character, changing occupation numbers, and
accurate momentum distributions. There are nevertheless
issues related to its convergence, efficiency and accuracy, that require
further study. The issues mainly concern the computational details of the classical trajectories, in particular, the large number of trajectories needed for convergence and also the issue of classical autoionization.

 Although a large number of
trajectories is required for convergence for our two-electron model
atoms, we expect that the ``forward-backward'' nature of the
semiclassical computation in Ref.~\cite{RRM10} for the two-body
reduced density-matrix will lead to favorable scaling with the number
of electrons. Because the prescription of Ref.~\cite{RRM10} only calls
for the correlation component of the dynamics to be treated
semiclassically, less accurate and more efficient semiclassical
methods, such as thawed Gaussian or even just quasiclassical
propagation (evolving an initial distribution just classically) may be
worth exploring: the aim is just to capture ``enough'' correlation in
a physical way to be used in conjunction with the exact treatment of 
the one-body terms in the equation of motion for $\rho(\br,\br',t)$.

How to deal with classical autoionization in a consistent and
practical way needs further study. When a strong field is present and
electrons are unbound at least for some time, the effect of classical
autoionization may be relatively small enough for times of interest
that a simple termination of trajectories when they get ``too far
away'' makes sense. However in other situations prescreening
trajectories by discarding at the start those which at any time reach
a given boundary, may be a better approach to avoid autoionizing
trajectories ``on their way out'' from distorting earlier
dynamics. Yet there is a delicate balance: if prescreening is done over
too long time, too many trajectories that are important at smaller
times get discarded. The solution of this problem also depends on what
is the quantity of interest, e.g. whether it is time-resolved
densities, or time-averaged spectra.

By including correlation semiclassically, the scheme of
Ref.~\cite{RRM10} is likely to provide more accurate dynamics than the
FG treatment of the entire dynamics shown here, as well as improving
over the TDHF and adiabatic TDDMFT results. The present results are
encouraging for this next step.

\section{Acknowledgements}
We thank Kenneth Kay and Rick Heller for very helpful discussions. 
Financial support from the National Science Foundation (CHE-0647913),
the Cottrell Scholar Program of Research Corporation and a grant of
computer time from the CUNY High Performance Computing Center under
NSF Grants CNS-0855217 and CNS-0958379, are gratefully acknowledged.

\end{document}